\documentclass[twocolumn,aps,a4paper,superscriptaddress]{revtex4-2}
\usepackage[T1]{fontenc}
\usepackage[utf8]{inputenc}
\usepackage{kpfonts}
\usepackage{amsmath}
\usepackage{amssymb}
\usepackage{times}
\usepackage{graphicx}
\usepackage[colorlinks,linkcolor=blue,citecolor=blue,urlcolor=blue,breaklinks]{hyperref}
\usepackage[normalem]{ulem}

\newcommand{\be}{\begin{equation}}
\newcommand{\ee}{\end{equation}}
\newcommand{\bea}{\begin{eqnarray}}
\newcommand{\eea}{\end{eqnarray}}

\let\polishl\l

\def\a{\alpha}
\def\b{\beta}

\def\d{\delta}

\def\e{\epsilon}

\def\k{\kappa}
\def\l{\lambda}

\def\p{\pi}

\def\r{\rho}

\def\t{\tau}

\def\F{\Phi}

\def\w{\omega}
\def\W{\Omega}

\def\Q{\Psi}

\def\blk{{\mathbf k}}

\def\bln{{\mathbf n}}

\def\blq{{\mathbf q}}

\def\blQ{{\mathbf Q}}

\def\callG{\mbox{$\mathcal{G}$}}

\def\callN{\mbox{$\mathcal{N}$}}
\def\callO{\mbox{$\mathcal{O}$}}

\def\iif{\infty}
\def\bra{\langle}
\def\ket{\rangle}

\def\Tr{{\rm Tr}}

\def\1op{\hat{\mathbbm{1}}}
\def\nn{\nonumber}

\def\bz{\mathbf 0}

\begin{document}

\preprint{APS/123-QED}

\title{Long-Lived Coherence between Incoherent Excitons revealed by 
Time-Resolved ARPES: An Exact Solution}

\author{Zhenlin Zhang}
\author{Wei Hu}
 \affiliation{
 Hefei National Research Center for Physical Sciences at the Microscale,\\ University of Science and Technology of China, Hefei, Anhui 230026, China}%Lines break automatically or can be forced with \\

\author{Enrico Perfetto}%
\affiliation{Dipartimento di Fisica, Universit{\`a} di Roma Tor Vergata, Via della Ricerca Scientifica 1,
00133 Rome, Italy}
\affiliation{INFN, Sezione di Roma Tor Vergata, Via della Ricerca Scientifica 1, 00133 Rome, Italy}

\author{Gianluca Stefanucci}%
\affiliation{Dipartimento di Fisica, Universit{\`a} di Roma Tor Vergata, Via della Ricerca Scientifica 1,
00133 Rome, Italy}
\affiliation{INFN, Sezione di Roma Tor Vergata, Via della Ricerca Scientifica 1, 00133 Rome, Italy}

\begin{abstract}
We investigate the exciton dynamics in an 
exactly solvable two-band model for semiconductors. The model 
incorporates light-matter, electron-electron and electron-phonon 
interactions, and captures exciton formation as well as the 
transition from the coherent to the incoherent regime.
We analyze excitonic polarization, populations and coherences, with 
special focus on their impact in Time-Resolved and Angle-Resolved 
Photoemission Spectroscopy (TR-ARPES). For nonresonant pumping with 
below-gap photon energies, TR-ARPES spectra 
reveal distinct excitonic replica and quantum beats 
persisting in the incoherent regime. These are due to a
coherence between different species of {\em incoherent} excitons. 
Such type of coherence is resistant to phonon dephasing, 
indicating that it follows different dynamics than those governing 
the coherences considered so far.                   
\end{abstract}

\maketitle

%\tableofcontents

\section{Introduction}
Experiments and theoretical advancements in Time-Resolved and 
Angle-Resolved Photoemission Spectroscopy (TR-ARPES) constitute a 
highly active field in condensed matter physics~\cite{boschini_time-resolved_2024}. This technique  
can probe electrons in occupied as well as unoccupied excited states. 
TR-ARPES has emerged as a powerful experimental 
tool in diverse areas, including the study of topological 
insulators~\cite{PhysRevLett.108.117403}, carrier 
dynamics~\cite{doi:10.1021/acs.nanolett.5b01967}, band structure 
control~\cite{doi:10.1021/acsnano.6b02622}, transient 
charge-density-wave (CDW) gap 
melting~\cite{doi:10.1126/science.1160778}, optical control of 
spin and valley polarization in excited 
carriers~\cite{PhysRevLett.117.277201}, and exciton 
dynamics~\cite{madeo_directly_2020,man_experimental_2021,dong_direct_2021}. 
It has also garnered 
significant interest from theorists aiming to provide accurate 
descriptions based on many-body 
theory~\cite{freericks_theoretical_2009,perfetto_first-principles_2016}.                    

Low-dimensional materials exhibiting excitonic effects demonstrate 
significant potential for optoelectronic device applications, 
including electrically driven light emitters, photovoltaic solar 
cells, photodetectors, and opto-valleytronic devices~\cite{mueller_exciton_2018}. 
Under quasi-resonant excitation, these systems generate a 
time-dependent 
polarization~\cite{haug2009quantum,ShaferWegenerbook,lindberg_effective_1988,koch_semiconductor_2006}, 
which can persist for a few hundred 
femtoseconds~\cite{moody_intrinsic_2015,jakubczyk_impact_2018,jakubczyk_coherence_2019}.    
In this coherent regime, the system is described by a quantum 
superposition of the ground state and exciton states, and we say that 
it contains {\em coherent} excitons.
After the pump has died off, phonon-assisted 
polarization-to-population transfer takes place, typically on a 
sub-picosecond timescale. This is the key mechanism governing the 
transition from the coherent to the incoherent    
regime~\cite{haug2009quantum,lindberg_effective_1988,thranhardt_quantum_2000,koch_semiconductor_2006,selig_exciton_2016,brem_exciton_2018,selig_dark_2018,sangalli_an-ab-initio_2018,stefanucci_excitonic_2025}.
In the incoherent regime, we do not have a quantum superposition, but 
rather an ensemble of ground-state and exciton states; therefore, we 
say that the system contains {\em incoherent} excitons.  
Phonons are thus initially responsible for converting coherent bright 
excitons into incoherent bright and dark  
excitons~\cite{zhang_experimental_2015,madeo_directly_2020}.
Subsequently, 
exciton-phonon~\cite{toyozawa_theory_1958,thranhardt_quantum_2000,jiang_exciton-phonon_2007,shree_observation_2018,chen_exciton-phonon_2020,cudazzo_first-principles_2020,antonius_theory_2022} 
and, at high excitation densities, exciton-exciton~\cite{sun_stimulated_2000,shahnazaryan_exciton-exciton_2017,erkensten_exciton-exciton_2021,perfetto_real_2022} 
interactions drive diffusion and, eventually, 
relaxation toward a thermal state.~\cite{chen_exciton-phonon_2020,chen_first-principles_2022,stefanucci_excitonic_2025}.

In semiconductors with Rydberg-like excitonic series, two main types 
of coherences can be distinguished.    
Ultrafast photoexcitations with below-gap photon-energies cause a 
transition from the ground state (G) $|\F_{g}\ket$ of energy $E_{g}$ to the quantum 
superposition state $|\Q\ket=\a|\F_{g}\ket+\sum_{\l}\b_{\l}|X_{\l}\ket$, where 
$|X_{\l}\ket$ are bright exciton (X) states  (one particle-hole over 
the ground-state) of energy $E_{\l}$. At clamped nuclei, the quantum 
average of an electronic observable $O(t)=\bra\Q(t)|\hat{O}|\Q(t)\ket$ 
generally oscillates in time with frequencies $(E_{\l}-E_{g})$ and 
$(E_{\l}-E_{\l'})$. 
The coherent regime refers to the quantum state of matter 
featuring G-X  coherence, leaving unspecified the X-X coherence. 
Of course, electron-phonon scattering is responsible for suppressing 
both types of coherence. However, what about the associated 
time scales? When coherent excitons, $|X\ket$,
are converted into incoherent excitons, $|X^{\rm inc}\ket$, the G-X 
coherence vanishes rather than transforming into 
a G-X$^{\rm inc}$ coherence. However, could   the 
X-X coherence~\cite{rustagi_coherent_2019,malakhov_exciton_2024} instead transform into a X$^{\rm inc}$-X$^{\rm inc}$ coherence 
that persists over longer time scales? If so, the mere absence of coherent 
excitons would not necessarily indicate the onset a quasi-stationary state.

TR-ARPES offers an 
invaluable tool to investigate these two different time scales, 
and to elucidate the nature of what is 
broadly defined as an ``exciton''.  In fact,  
excitonic structures within the gap 
can exhibit changes in shape, position, and intensity, indicating 
transitions between excited states with distinct excitonic 
features. While the incoherent regime has been investigated 
both 
theoretically~\cite{perfetto_first-principles_2016,steinhoff2017exciton,rustagi_photoemission_2018,christiansen_theory_2019} 
and 
experimentally~\cite{zhu_photoemission_2015,tanimura_dynamics_2019,madeo_directly_2020,dendzik_observation_2020,man_experimental_2021,wallauer_momentum-resolved_2021,fukutani_detecting_2021,dong_direct_2021}, 
the coherent regime 
has thus far been primarily explored through theoretical 
frameworks~\cite{perfetto_pump-driven_2019,rustagi_coherent_2019,perfetto_time-resolved_2020,perfetto_self-consistent_2020,perfetto_floquet_2020,chan_giant_2023}
(coherence times in the range of a 
few hundreds of femtoseconds or shorter make experiments challenging). 
As for the coherent-to-incoherent crossover, the focus of this work,
TR-ARPES remains a "hic sunt leones" territory.

The coherent-to-incoherent crossover is typically addressed using the 
excitonic Bloch equations, which is a set of coupled equations for the
{\em populations} of coherent and incoherent excitons~\cite{haug2009quantum,lindberg_effective_1988,thranhardt_quantum_2000,koch_semiconductor_2006,selig_exciton_2016,brem_exciton_2018,selig_dark_2018,stefanucci_excitonic_2025}. 
However, 
populations alone are not sufficient to extract TR-ARPES spectra. 
Moreover, these equations neglect X-X coherences from the outset.  

In this work we resort to the exact solution of a two-band model with 
electron-electron and electron-phonon interactions to correlate  
G-X and X-X coherences with TR-ARPES signatures. Our analysis 
suggests that phonons suppress the G-X coherence but transform the X-X 
coherence into a robust X$^{\rm inc}$-X$^{\rm inc}$ coherence, which 
survives on much longer timescales.
Although the two-band model ignores the atomistic details 
of real 
excitonic materials, our results indicate that the X$^{\rm inc}$-X$^{\rm inc}$  
coherence is significantly more resilient than the G-X
one. This effect points to the potential of creating exciton-driven 
Floquet matter as an alternative approach to Floquet physics in 
material science.

The plan of the paper is as follows. 
We discuss the model Hamiltonian in Section~\ref{modelsec}. In 
Section~\ref{exactsec} we derive 
the exact time-dependent many-body state along with other quantities 
like the one-particle Green's function, the exciton Green's function 
and the many-body reduced density matrix. The time evolution of the 
system is investigated numerically in Section~\ref{resultsec} for 
resonant and nonresonant photoexcitations. Finally, we draw our 
conclusions in Section~\ref{concsec}.

\section{Model Hamiltonian}
\label{modelsec}

We consider the  two-band  model Hamiltonian introduced in 
Ref.~\cite{stefanucci_from-carriers_2021}:         
\begin{equation}\label{total hamiltonian}
    \hat{H}=\hat{H}_\text{e}+\hat{H}_\text{ph}+
	\hat{H}_\text{e-ph}.
\end{equation}
The first term is the pure electronic Hamiltonian,
\begin{equation}
	\hat{H}_\text{e}=\sum_\blk(\epsilon^c \hat{c}_\blk^{\dagger} 
	\hat{c}_\blk+\epsilon_\blk^v \hat{v}_\blk^{\dagger} \hat{v}_\blk)-
	\frac{1}{\mathcal{N}}\sum_{\blk \blk_1 \blk_2} \!U_\blk\; 
	\hat{c}_{\blk_1+\blk}^{\dagger} \hat{v}_{\blk_2+\blk} 
	\hat{v}_{\blk_2}^{\dagger} \hat{c}_{\blk_1},
\end{equation}
characterized by a {\em flat} conduction band and a dispersive valence 
band, $U_\blk$ is the Coulomb interaction between 
conduction-band electrons and valence-band holes, and  
$\mathcal{N}$ denotes the number of discrete crystal momenta in the 
first Brillouin zone. 
The second term is the Hamiltonian of a single phonon mode
\begin{align}
\hat{H}_\text{ph}=
\sum_\blq \omega_\blq \hat{b}_\blq^{\dagger} \hat{b}_\blq, 
\end{align}
while 
\begin{equation}
    \hat{H}_\text{e-ph}=\frac{1}{\sqrt{\mathcal{N}}} 
    \sum_{\blk \blq} (g_\blq
    \hat{c}_{\blk+\blq}^{\dagger} \hat{c}_\blk \hat{b}_\blq+
    \text{H.c.})
\end{equation}
describes the electron-phonon interaction, characterized by a 
$\blk$-independent coupling. The flatness of the conduction band and 
the absence of electron-phonon interactions for electrons in the valence 
band enable  an analytic solution for the time-dependent 
many-body state, as discussed below. Despite these simplifications, 
the model correctly captures the formation of coherent excitons, the 
 phonon-induced suppression of G-X 
coherences and the formation of 
incoherent excitons~\cite{stefanucci_from-carriers_2021}.

To describe Rydberg-like excitonic series~\cite{chernikov_exciton_2014,he_tightly_2014}
a Yukawa-screened Coulomb interaction is used, $U_\blk=4\pi 
u/(|\blk|^2+\k^2)$, where $u$ is the coupling strength and $\k$ is 
the inverse of the screening length.
For relatively weak Coulomb interaction 
compared to the band gap, the ground state $|\Phi_g\rangle$ consists 
of a completely filled valence band and an empty conduction band; 
we set its energy $E_{g}=0$. In the subspace of one 
particle-hole states with total momentum $\blQ$, 
the electronic Hamiltonian can be diagonalized.
The eigenvalue problem is the well known Bethe-Salpeter equation
\begin{align}
(\e^{c}-\e^{v}_{\blk})A^{\l\blQ}_{\blk}-\frac{1}{\callN}
\sum_{\blk'}U_{\blk-\blk'}A^{\l\blQ}_{\blk'}=E_{\l\blQ}A^{\l\blQ}_{\blk},
\end{align}
yielding bound excitonic states and unbound 
electron-hole pair states. In the following, we refer to all types of 
eigenstates as excitons. They can be written as
\begin{align}
|X_{\l\blQ}\rangle = \hat{X}^\dagger_{\l\blQ}|\Phi_g\rangle = 
\sum_\blk A^{\l\blQ}_{\blk} 
\hat{c}^\dagger_{\blk+\blQ}\hat{v}_\blk|\Phi_g\rangle.
\label{fqex}
\end{align}
Due to the flatness of the conduction 
band, both the wavefunctions and energies are independent of 
$\blQ$:
\begin{align}
A^{\l\blQ}_{\blk}&=A^{\l}_{\blk},
\label{xwf}
\\
E_{\l\blQ}&=E_{\l}.
\label{xene}
\end{align}

The exciton dynamics is initiated by an external pump field. The  light-matter coupling
is described as:
\begin{equation}
    \hat{H}_\text{light-matter}(t)=\sum_{\blk} e(t)d_\blk\hat{c}_{\blk}^{\dagger} \hat{v}_\blk +\text{H.c.},
\end{equation}
where $d_\blk$ is the dipole moment and the electric field of the pump
\begin{align}
e(t)=e_{0}\sin^2\left( 
\frac{\pi(t-T_P)}{T_P}\right)\sin[\omega_P(t-T_P)]
\end{align}
is active from $t=-T_P$ to $t=0$. 
The pump frequency $\omega_P$ will be tuned either to resonate with 
an exciton or set off-resonance, but still below the gap. 
In this work, 
radiative recombination is neglected, hence the total charge in each 
band is a conserved quantity for positive times.

For simplicity, we assume
that electron-phonon scattering is negligible during the 
pump. This assumption is justified provided that the time-scale for phonon 
generation is longer than $T_{P}$ (or equivalently, the dynamics 
during the pump is weakly affected by phonons). 
Then, for $-T_{P}<t<0$, only bright excitons are 
generated, and for weak pumps (linear regime) 
the many-body state of the system reads
\begin{align}
|\Psi(t)\rangle
=&\alpha(t)|\Phi_g\rangle
+\sum_\l
\b_\l(t)
|X_{\l\bz}\rangle.
\end{align}
The time-dependent coefficients $\alpha(t)$ and $\b_{\l}(t)$ can 
easily be obtained numerically by solving the system of linear 
differential equations:
\begin{equation}
\begin{aligned}
i\dot{\alpha}(t)&=\sum_\l\beta_\l(t)\Omega^{\ast}_\l(t),\\
i\dot{\beta}_\l(t)&=E_{\l}\b_{\l}(t)+\alpha(t)\Omega_\l(t),
\end{aligned}
\end{equation}
where
\begin{equation}
\begin{aligned}
\Omega_\l(t)=\sum_\blk e(t)d_\blk A_\blk^{\l\ast}.
\end{aligned}
\end{equation}

\section{Exact Solution}
\label{exactsec}

The analytic expression of the time-dependent many-body state of the system for positive 
times is the first main result of this work. In Appendix~\ref{aapp} we show 
that       
\begin{align}
|\Psi(t>0)\rangle
&=\alpha(0)|\Phi_g\rangle
+\sum_\l
\b_\l(0)\ell(t)e^{-iE_\l t}
\nn\\
&\times\left(
|X_{\l\bz}\rangle+\sum_{\blQ}\sqrt{S_{\blQ}(t)}|X_{\l \blQ}^{\text{inc}}(t)\rangle\right).
\label{exact_wf}
\end{align}
Let us examine the physical implications of this result.
In Eq.~(\ref{exact_wf}), the {\em 
many-body states} $|X_{\l \blQ}^{\text{inc}}(t)\rangle$ 
of the incoherent $\l \blQ$-excitons, 
which have so far been characterized solely in terms of populations 
$N^{\rm inc}_{\l\blQ}$,   
appear.
 The explicit form of the states is
\begin{align}
|X_{\l \blQ}^{\text{inc}}(t)\rangle
=&\frac{1}{\sqrt{S_\blQ(t)}}
\sum_{M=1}^\infty
\frac{1}{M!}
\sum_{ \blq_1 \ldots \blq_M}
\delta_{\blq_1\ldots +\blq_M,-\blQ}
\nn \\
&\times f_{\blq_1}(t) \ldots f_{\blq_M}(t)
\hat{b}^{\dag}_{\blq_1}\ldots\hat{b}^{\dag}_{\blq_M}
|X_{\l\blQ}\rangle,
\label{incex}
\end{align}
where
\begin{align}
f_\blq(t)=\frac{1}{\sqrt{\mathcal{N}}} 
\frac{g_\blq}{\omega_\blq}\left(e^{-i \omega_\blq t}-1\right)
\end{align}
is the Langreth function~\cite{langreth_singularities_1970}, and the 
scattering function 
\begin{align}
S_\blQ(t)=
\frac{1}{\mathcal{N}}
\sum_\bln e^{i\blQ\cdot\bln}
\exp{\left[
\sum_\blq 
|f_\blq(t)|^{2}
e^{-i\blq\cdot\bln}
\right]} 
-\delta_{\blQ,\bz},
\label{scattfunc}
\end{align}
is defined in terms of a sum  over all unit cell vectors 
$\bln$ of the crystal. In Appendix~\ref{aapp} we show that $S_\blQ(t)$ is real 
and strictly positive for all times. According to Eq.~(\ref{incex}), 
the incoherent exciton state is  an 
exciton dressed by a cloud of phonons, more simply referred to as 
an exciton-polaron. Notice that the total momentum of 
$|X_{\l \blQ}^{\text{inc}}(t)\rangle$
 is zero; the label $\blQ$ refers to the 
momentum of the pure electronic part $|X_{\l\blQ}\rangle$, see 
Eq.~(\ref{fqex}).

The (bright) coherent, $|X_{\l\bz}\rangle$, and (momentum-dark) incoherent, 
$|X_{\l \blQ}^{\text{inc}}(t)\rangle$, exciton states are mutually 
orthonormal, see Appendix~\ref{aapp},
\begin{subequations}
\begin{align}
\bra X_{\l\bz}|X_{\l'\bz}\ket&=\d_{\l\l'},
\\
\bra X_{\l \blQ}^{\text{inc}}(t)|X_{\l' 
\blQ'}^{\text{inc}}(t)\ket&=\d_{\l\l'}\d_{\blQ,\blQ'},
\\
\bra X_{\l\bz}|X_{\l' 
\blQ'}^{\text{inc}}(t)\ket&=0.
\end{align}
\label{orthrel}
\end{subequations}
Therefore, the unitary evolution implies that 
\begin{align}
\left|\b_{\l}(0)\ell(t)\right|^{2}\left(
1+\sum_{\blQ}S_{\blQ}(t)\right)=1-|\a(0)|^{2}.
\label{unitev}
\end{align}
The function
\begin{align}
\ell(t)=\exp \left[\frac{1}{\mathcal{N}} 
\sum_\blq\left(\frac{g_\blq}{\omega_\blq}\right)^2\left(-1+e^{-i 
\omega_\blq t}+i \omega_\blq t\right)\right],   
\end{align}
is a decaying function of time, see Section~\ref{resultsec}, and 
provides the time scale of the G-X coherence, see again Eq.~(\ref{exact_wf}).
The scattering function $S_\blQ(t)$ compensates the decay of 
$\ell(t)$ in such a way that the product $\ell(t)\sqrt{S_\blQ(t)}$ 
remains finite in the long time limit, see Eq.~(\ref{unitev}). 
Thus, Eq.~(\ref{exact_wf}) conveys a clear physical message: electron-phonon 
scattering causes a depopulation of coherent excitons, which is 
compensated by an increase of incoherent  
excitons with finite momentum and finite phonon numbers ($M > 0$).

\subsection{Exciton states and exciton populations}

Our definition of incoherent exciton states is fully consistent 
with the definitions of coherent and incoherent exciton populations. The 
total number of excitons is typically written 
as~\cite{kira_many-body_2006,stefanucci_excitonic_2025}
\begin{align}
N_{\l\blQ}\left(t\right) =&\bra\Q(t)|
\hat{X}_{\lambda \blQ}^{\dagger}  \hat{X}_{\lambda \blQ}|\Q(t)\ket
=\delta_{\blQ,\bz}|X_{\lambda \bz}(t)|^{2}
+N^{\text{inc}}_{\lambda \blQ}\left(t\right),
\label{totaln}
\end{align}
where 
\begin{align}
X_{\lambda \bz}(t)\equiv \bra\Q(t)|\hat{X}_{\lambda \bz}|\Q(t)\ket.
\end{align}
The quantity $X_{\lambda \bz}(t)$
is referred to as the excitonic polarization, and its square modulus yields 
the number of coherent $\l$-excitons. From 
Eq.~(\ref{exact_wf}) we find
\begin{align}
X_{\lambda \bz}(t) 
& =\alpha^*(0)
\sum_{\l'}
\b_{\l'}(0)\ell(t)e^{-iE_{\l'} t}
\left\langle\Phi_{g}\right|
\hat{X}_{\lambda \bz}  
\left|X_{\l'\bz}\right\rangle\nn \\
& =\alpha^*(0)
\b_{\l}(0)\ell(t)e^{-iE_{\l} t},
\label{xlcoh}
\end{align}
according to which the number of coherent $\l$-excitons is
\begin{align}
N^{\rm coh}_{\l}(t)=|X_{\lambda \bz}(t)|^{2}=
|\b_{\l}(0)\ell(t)|^{2}+\callO(1-|\a(0)|^{2}).
\label{ncoh}
\end{align}
(We recall that in the linear response regime $|\a(0)|^{2}$ remains 
close to unity).

Let us now introduce the projection operators onto the space of coherent 
and incoherent excitons:
\begin{align}
\hat{P}^{\rm coh}\equiv \sum_{\l}|X_{\l\bz}\ket\bra X_{\l\bz}|,
\\
\hat{P}^{\rm inc}(t)\equiv \sum_{\l\blQ}|X^{\rm inc}_{\l\blQ}(t)\ket\bra X^{\rm 
inc}_{\l\blQ}(t)|.
\end{align}
Taking into account the orthonormality relations in  
Eq.~(\ref{orthrel}) we have
\begin{align}
|\Q(t)=\a(0)|\F_{g}\ket+\hat{P}^{\rm coh}|\Q(t)\ket+	
\hat{P}^{\rm inc}(t)|\Q(t)\ket.
\label{projstate}
\end{align}
It is straightforward to verify that the value of $N^{\rm 
coh}_{\l}(t)$ in Eq.~(\ref{ncoh})
is the same as the exciton number operator averaged over the 
coherent component of the 
many-body state, that is
\begin{align}
\bra\Q(t)|\hat{P}^{\rm coh}
\hat{X}_{\lambda \blQ}^{\dagger}  \hat{X}_{\lambda \blQ}
\hat{P}^{\rm coh}|\Q(t)\ket=\d_{\blQ,\bz}|\b_{\l}(0)\ell(t)|^{2}.
\label{cohexnum}
\end{align}
Additionally, since the electron-phonon interaction preserves the 
total number of electrons in each band,  we can rewrite Eq.~(\ref{totaln}) 
as
\begin{align}
N_{\l\blQ}(t)&=\bra\Q(t)|\hat{P}^{\rm coh}
\hat{X}_{\lambda \blQ}^{\dagger}  \hat{X}_{\lambda \blQ}
\hat{P}^{\rm coh}|\Q(t)\ket
\nn\\
&+
\bra\Q(t)|\hat{P}^{\rm inc}(t)
\hat{X}_{\lambda \blQ}^{\dagger}  \hat{X}_{\lambda \blQ}
\hat{P}^{\rm inc}(t)|\Q(t)\ket.
\label{}
\end{align}
The first term in this equation
coincides with the first term of Eq.~(\ref{totaln}), see 
Eqs.~(\ref{cohexnum}) and~(\ref{ncoh}), and therefore
\begin{align}
N^{\rm inc}_{\l\blQ}(t)&=\bra\Q(t)|\hat{P}^{\rm inc}(t)
\hat{X}_{\lambda \blQ}^{\dagger}  \hat{X}_{\lambda \blQ}
\hat{P}^{\rm inc}(t)|\Q(t)\ket
\nn\\
&=|\b_{\l}(0)\ell(t)|^{2}S_{\blQ}(t).
\label{ninc}
\end{align}

It is worth noticing that Eqs.~(\ref{ncoh}) and (\ref{ninc}) together 
with Eq.~(\ref{unitev}) imply that the total number of excitons 
remain constant after the pump (positive times):
\begin{align}
\sum_{\l}\left(N^{\rm coh}_{\l}(t)+\sum_{\blQ}N^{\rm 
inc}_{\l\blQ}(t)\right)={\rm const},\quad t>0.
\label{conslaw}
\end{align}
	
\subsection{One-particle Green's Function}

To calculate the TR-ARPES spectrum we need the one-particle lesser 
Green's function, which is defined as
\begin{align}
G_{cc\blk}(t, t^{\prime}) =
&i\left\langle
\Psi(t^{\prime})
\right| 
\hat{c}_\blk^{\dagger} 
e^{-i \hat{H}(  t^{\prime}-t)}
\hat{c}_\blk
|\Psi(t)\rangle .
\label{GF1}
\end{align}
Writing the many-body state as in Eq.~(\ref{projstate}) we see that 
the cross product between coherent and incoherent states vanish, and 
therefore
\begin{align}
G_{cc\blk}(t, t^{\prime})
=G_{cc\blk}^{\text{coh}}(t, t^{\prime})
+G_{cc\blk}^{\text{inc}}(t, t^{\prime}) ,
\label{GF2}
\end{align}
where 
\begin{align}
G_{cc\blk}^{\text{coh}}
(t, t^{\prime})=&i\bra \Psi(t^{\prime})|\hat{P}^{\rm coh}
\hat{c}_\blk^{\dagger} 
e^{-i \hat{H}(  t^{\prime}-t)}
\hat{c}_\blk\hat{P}^{\rm coh}
|\Psi(t)\rangle 
\nn\\
=&i
\sum_{\l\l'}\b_{\l}(t)\b^{\ast}_{\l'}(t')
\ell(t)\ell^{\ast}(t')
A_\blk^{\l} A^{\l'\ast}_\blk\nn \\
&\times e^{-i(\epsilon_\blk^v+E_\l) t}
e^{i(\epsilon_\blk^v+E_{\l'})t'},
\label{gcccoh}
\end{align}
and
\begin{align}
G_{cc\blk}^{\text{inc}}
(t, t^{\prime})=&i\bra \Psi(t^{\prime})|\hat{P}^{\rm inc}(t')
\hat{c}_\blk^{\dagger} 
e^{-i \hat{H}(  t^{\prime}-t)}
\hat{c}_\blk\hat{P}^{\rm inc}(t)
|\Psi(t)\rangle 
\nn\\
=&\sum_\blQ
S_\blQ(t,t')
G_{cc\blk+\blQ}^{\text{coh}}
(t, t^{\prime}).
\label{gccinc}
\end{align}
The function
\begin{align}
&S_\blQ(t,t')=
\frac{1}{\mathcal{N}}
\sum_\bln e^{i\blQ \cdot\bln}
\nn \\
&\quad\times\exp{\left[
\sum_\blq 
f^*_\blq(t') f_\blq(t)
e^{-i\omega_\blq(t'-t)-i\blq\cdot\bln}
\right]} 
-\delta_{\blQ,0},
\end{align}
extends Eq.~(\ref{scattfunc}) away from the time diagonal since
\begin{align}
S_\blQ(t,t)=S_\blQ(t).
\end{align}
The derivation of these results can be found in Appendix~\ref{bapp}.

For completeness we also report the off-diagonal lesser Green's function 
\begin{align}
G_{cv\blk}\left(t, t^{\prime}\right) & =i\bra\Psi(t^{\prime})|
\hat{v}_\blk^{\dagger} e^{-i \hat{H}\left(t^{\prime}-t\right)} 
\hat{c}_\blk|\Psi(t)\ket \nn\\
&=i \alpha^*(t') \sum_\l 
\b_{\l}(t)\ell(t) A^\l_\blk e^{-iE_\l t} e^{i 
\epsilon_\blk^v\left(t^{\prime}-t\right)}.
\label{gcv}
\end{align}

Notice that the expressions for the Green's functions in 
Eqs.~(\ref{gcccoh},\ref{gccinc},\ref{gcv}) are valid for all $t$ and 
$t'$ provided that we extend the definition of $\a(t)$ and $\b(t)$ to 
positive times as
\begin{align}
\a(t>0)=\a(0),\quad\b(t>0)=\b(0),
\end{align}
and the definition of $f_{\blq}(t)$ and $\ell(t)$ at negative times as
\begin{align}
f_{\blq}(t<0)=0,\quad\ell(t<0)=1.	
\end{align}

In Section~\ref{resultsec} we use the Green's function to calculate 
the TR-ARPES signal at pump-probe delay $T$ by Fourier transforming 
$G_\blk^{cc}(t, t')$ with respect to the relative time $\tau$
\begin{align}
A_\blk(T, \omega)=-i \int d \tau e^{i \omega \tau} G_{cc\blk}
\left(T+\frac{\tau}{2}, T-\frac{\tau}{2}\right).  
\label{spectralf}
\end{align}	
The temporal evolution of the spectral function from the overlapping 
regime to the nonoverlapping incoherent regime is the second main 
result of this work.

We also use the equal-time Green's function to calculate the 
momentum-resolved occupations in the conduction band
\begin{align}
n_{c\blk}(t)
=&-iG_{cc\blk}(t,t)=n_{c\blk}^{\text{coh}}(t)+n_{c\blk}^{\text{inc}}(t),
\label{nckanalyt}
\end{align}
and the total polarization
\begin{align}
P(t)  =-\frac{i}{\mathcal{N}} \sum_\blk d_\blk G_{cv\blk}(t, t)+\text { H.c. }
\end{align}

\subsection{Excitonic Green's function}
From the exact many-body state we can also calculate the excitonic 
Green's function, which is defined according 
to~\cite{stefanucci_excitonic_2025}
\begin{align}
N_{\lambda\lambda'\blQ}
(t, t^{\prime})  
=&\left\langle\Psi(t^{\prime})\right|
\hat{X}_{\lambda' \blQ}^{\dagger} 
e^{-i \hat{H}( t^{\prime}-t)} 
\hat{X}_{\lambda \blQ}
|\Psi(t)\rangle \nn \\
=&\delta_{\blQ,\bz}
N^{\text{coh}}_{\lambda\lambda'}
(t, t^{\prime})
+N_{\lambda\lambda' \blQ}^{\text{inc}}
(t, t^{\prime}).
\label{X_GF1}
\end{align}
In Appendix~\ref{bapp} we show that
\begin{align}
N^{\text{coh}}_{\lambda\lambda'}
(t, t^{\prime})
&=\b_{\l}(t)\b^{\ast}_{\l'}(t')
\ell(t)\ell^{\ast}(t') e^{-iE_\lambda t+iE_{\lambda'} t'} 
\nn\\
&=X_{\l\bz}(t)X^{\ast}_{\l'\bz}(t),
\end{align}
and 
\begin{align}
N_{\lambda\lambda'\blQ}^{\text{inc}}
(t, t^{\prime})
=S_{\blQ}(t,t')
N^{\text{coh}}_{\lambda\lambda'}
(t, t^{\prime}).
\label{inctocoh}
\end{align}

The equal-time values of these Green's functions yield the coherent 
exciton density matrix 
\begin{align}
N^{\text{coh}}_{\lambda\lambda'}
(t)\equiv N^{\text{coh}}_{\lambda\lambda'}
(t, t),
\label{Xcoh_dm}
\end{align}
and the incoherent exciton density matrix
\begin{align}
N^{\text{inc}}_{\lambda\lambda'\blQ}
(t)\equiv N^{\text{inc}}_{\lambda\lambda'\blQ}
(t, t).
\label{Xinc_dm}
\end{align}
The diagonal values ($\l=\l'$) of the density matrices correctly 
reduce to the coherent and incoherent exciton populations of 
Eqs.~(\ref{ncoh}) and (\ref{ninc}).

\subsection{Many-Body Reduced Density Matrix}

To characterized the electronic subsystem at a certain time we also 
calculate the  many-body reduced density matrix, obtained by tracing 
the full many-body density matrix over the phononic degrees of freedom 
\begin{align}
\hat{\rho}_{\mathrm{el}}(t)=\operatorname{Tr}_{\mathrm{ph}}
\{|\Psi(t)\rangle\langle\Psi(t)|\}.
\end{align}
In Appendix~\ref{bapp} we show that
\begin{align}
\hat{\rho}_{\mathrm{el}}(t) &=|\alpha(t)|^2
 |\Phi_g\ket
 \bra\Phi_g|+ \sum_{\lambda\lambda'}N^{\text{coh}}_{\lambda\lambda'}(t)
 |X_{\l'\bz}\ket\bra X_{\l\bz}|\nn\\
 &+\sum_\lambda\left(
 X_{\lambda \bz}(t)
|X_{\l\bz}\ket\bra
\Phi_g|+X_{\lambda \bz}^{\ast}(t)
|\Phi_g\ket\bra
X_{\l\bz}|
\right)
 \nn\\
& 
+\sum_{\lambda\lambda'\blQ}N_{\lambda\lambda'\blQ}^{\text{inc}}(t)
|X_{\lambda'\blQ}\ket\bra X_{\lambda\blQ}|
\label{eldm}
\end{align}
Using Eq.~(\ref{unitev}) we can easily check that $\Tr_{\rm 
el}[\hat{\rho}_{\mathrm{el}}(t)]=1$ for all times, as it should.

Let us discuss Eq.~(\ref{eldm}). Immediately after pumping
the electronic subsystem exhibits G-X and X-X coherences, with 
$X_{\lambda \bz}(t)$ governing the timescale over which this coherent regime 
persists. The X-X coherence  decays twice as fast since 
$N^{\text{coh}}_{\lambda\lambda'}=X_{\lambda \bz}X_{\lambda' 
\bz}^{\ast}$. As time passes, both coherent exciton populations and 
X-X coherences are 
transferred to incoherent excitons. In the  incoherent regime, we report a
long-lasting X$^{\rm inc}$-X$^{\rm inc}$ coherence. 
% The long-lasting coherence  is a direct consequence of the exact 
% solution. 
From Eqs.~(\ref{inctocoh}) and (\ref{xlcoh}) we have 
\begin{align}
N_{\lambda
\lambda'\blQ}^{\text{inc}}(t)&=	S_{\blQ}(t)X_{\lambda \bz}(t)X_{\lambda' 
\bz}^{\ast}(t)
\nn\\
&=\frac{\b_{\l'}^{\ast}(0)}{\b^{\ast}_{\l}(0)}
N^{\text{inc}}_{\lambda\blQ}(t)e^{-i(E_{\l}-E_{\l'})t}.
\label{Nllp}
\end{align}
Taking into account that the total number of 
excitons remains conserved, see Eq.~(\ref{conslaw}), we see that 
the X$^{\rm inc}$-X$^{\rm inc}$
coherence  does not vanish as time 
increases. While this conclusion depends on the chosen model 
Hamiltonian, it suggests that the X$^{\rm inc}$-X$^{\rm inc}$ coherence
can be far more robust than one might expect.

\section{Results}
\label{resultsec}

\begin{figure}[tbp]
\includegraphics[width=0.45\textwidth]{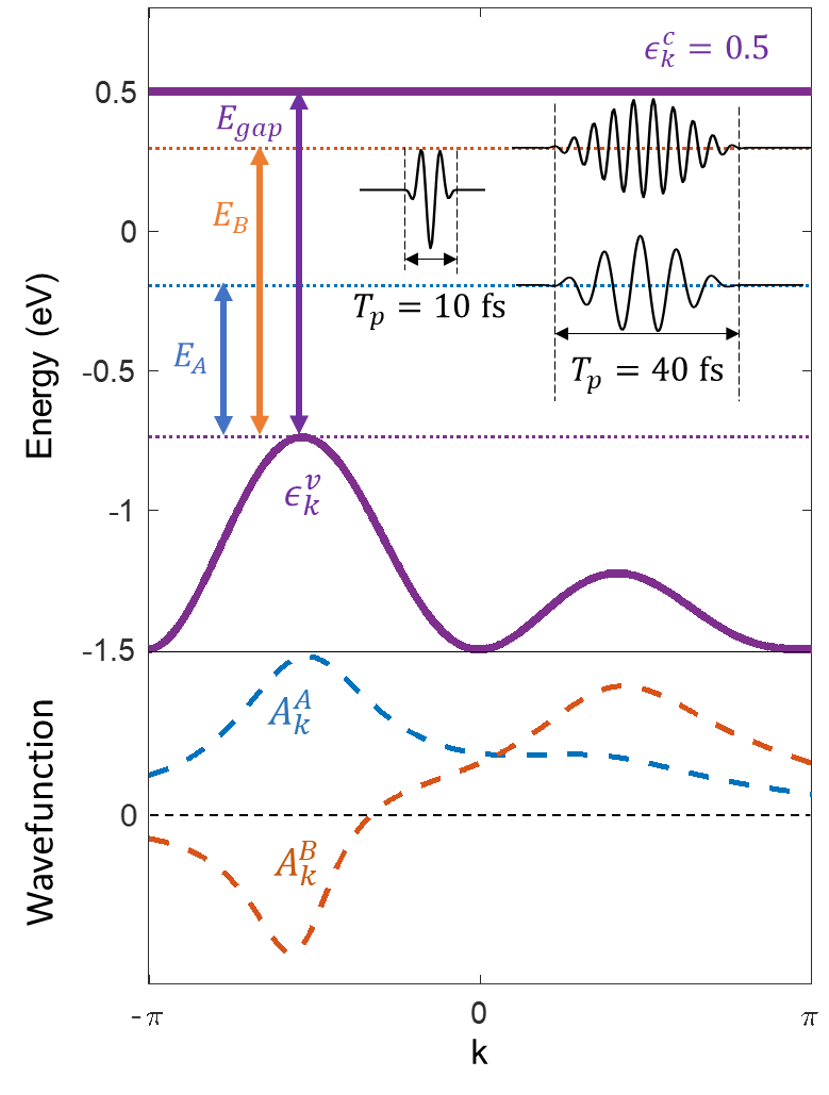}
\caption{(Top) Band structure and pump profiles. The arrows indicate 
the energy of the A and B excitons and the gap. For resonant 
excitations $\omega_P=E_A=0.54$~eV or $\omega_P=E_B=1.03$~eV whereas for nonresonant 
excitations $\omega_P=0.99$~eV.
(Bottom)
Excitonic wavefunctions in momentum space.}   
\label{fig:paremeters}
\end{figure}

We consider a one-dimensional system with a flat conduction band, 
$\epsilon^c_{k}=\e^{c}=0.5$~eV, and a dispersive valence band, 
$\epsilon_k^v=\big([1-k/(2\pi)]\cos^2{|k+\pi/2|}-1.5\big)$~eV, where 
the dimensionless quasi-momentum $k\in(-\p,\p)$.
The band structure is illustrated in Fig.~\ref{fig:paremeters}(top). 
The valence band has two local maxima, the absolute maximum 
(VBM) being at energy $-0.74$~eV, and the band gap is
$E_\text{gap}=1.24$~eV. For the statically screened Coulomb 
interaction we take $U_k=0.3\times{4\pi}/{(k^2+1.2^2)}$~eV, giving 
rise to two bound exciton states of energy $E_A\simeq 0.54$~eV  and 
$E_B\simeq 1.03$~eV. 
The corresponding exciton wavefunctions are 
shown in Fig.~\ref{fig:paremeters}(bottom) and resemble the first two eigenstates of a 1D 
infinite potential well.  
As for the phonon part, we consider acoustic phonons with dispersion 
$\omega_q=0.45 \sin|q|$~eV and an 
electron-phonon coupling  $g_q=0.1(\cos q+1)$~eV.

The system is driven out of equilibrium by a pump field with  
$k$-independent Rabi frequencies $\W_{k}=e_{0}d_{k}=10^{-5}$~eV 
(linear response regime). We study two different pump profiles, as  illustrated in 
Fig.~\ref{fig:paremeters}(top). 
All calculations 
employ $\mathcal{N}=280$ $k$-points to ensure sufficient resolution 
in momentum space.

\subsection{Resonant Pumping}

\begin{figure}[tbp]
\includegraphics[width=0.48\textwidth]{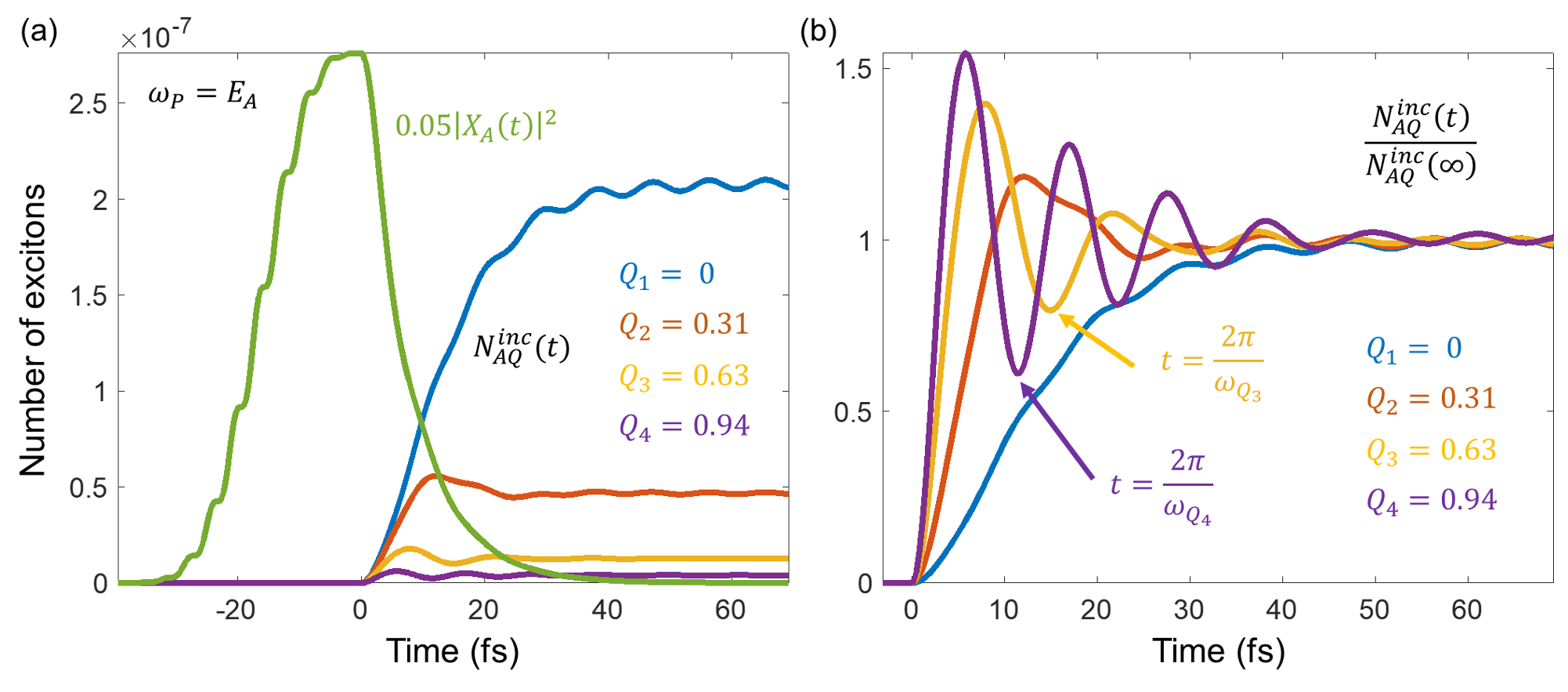}
\caption{\label{fig:Nx} (a) Population of coherent phonons $|X_{A}(t)|^{2}$ and 
population of incoherent excitons $N^\text{inc}_{AQ}(t)$ versus time for 
resonant excitation at frequency 
$\omega_P=E_{A}=0.54$~eV. (b) Population of incoherent excitons 
normalized to the asymptotic value.}    
\end{figure}

We tune the pump frequency $\omega_P=E_{A}$ in resonance with the 
A-exciton and set the pump duration $T_{P}\simeq 40$~fs (about 
five excitonic periods). In 
Fig.~\ref{fig:Nx}(a) we report the population of coherent excitons  
$|X_{A 0}(t)|^{2}$ ($X_{B0}(t)$ is vanishingly small) and the population of 
incoherent excitons for a few representative $Q$-points.
For resonant pumping no X-X coherence exists, and hence no X$^{\rm 
inc}$-X$^{\rm inc}$ coherence can develop.
The excitonic polarization increases 
monotonically (the superimposed oscillations have the same frequency 
as the pump pulse) until the end of the pump ($t=0$) and then decreases 
exponentially due to phonon-induced decoherence: $|X_{A 0}(t)|\simeq e^{-t/\t_{\rm 
decoh}}$. The decoherence 
time can be calculated independently~\cite{stefanucci_from-carriers_2021} and is given by
\begin{align}
\t_{\rm 
decoh}=\frac{2}{g_{0}^{2}}\left.\frac{d\w_{q}}{dq}\right|_{q=0}
\simeq 14.8\;{\rm fs}.
\label{decohtime}
\end{align}
We observe that the excitonic Bloch equations~\cite{haug2009quantum,lindberg_effective_1988,thranhardt_quantum_2000,koch_semiconductor_2006,selig_exciton_2016,brem_exciton_2018,selig_dark_2018,stefanucci_excitonic_2025} 
provide a formula for  
the decoherence time in terms of the so called {\em exciton-phonon 
coupling}. 
In our two-band model, the exciton-phonon 
coupling for the scattering $X_{\l Q}\to X_{\l'Q'}$ reads
\begin{align}
\callG^{\l\l'}(Q,Q')=\sum_{k}A^{\l Q\ast}_{k}g_{Q'-Q}A^{\l' Q'}_{k}	
=\d_{\l\l'}g_{Q'-Q},
\label{exphoncoup}
\end{align}
where in the last equality we use that $A^{\l Q}_{k}$ is 
independent of $Q$, see Eq.~(\ref{xwf}), and that the exciton 
wavefunctions are orthonormal. This explains the appearance of $g$ 
instead of $\callG$ in Eq.~(\ref{decohtime}).
We also emphasize that $1/\t_{\rm decoh}$ is quadratic in the 
electron-phonon coupling. Thus, higher order contribution 
vanish identically for this model Hamiltonian. 

The decrease of coherent excitons is concomitant with the increase of 
incoherent (momentum-dark) A-excitons (no B-excitons are generated). 
The rate of growth depends on 
the center-of-mass momentum $Q$, and is accompanied by superimposed 
oscillations of period $2\p/\w_{Q}$, see Fig.~\ref{fig:Nx}(b).
For $t\gtrsim 66$~fs, $X_{A 0}(t)$ vanishes, and 
the system enters the incoherent regime, characterized by the 
complete absence 
of G-X coherence. 
In this regime, only incoherent A-excitons exist. Their populations 
cease to oscillate and attain a steady value.
Interestingly, the steady-state populations are peaked at 
$Q=0$, despite the fact that the exciton energies are 
independent of $Q$, see Eq.~(\ref{xene}). Although the model investigated is an 
oversimplification of real excitonic materials, this result 
highlights that excitonic populations described by a Bose distribution 
(as predicted by the excitonic Bloch equations) is an approximation.
In fact,  
the extent to which excitons can be approximated as 
bosons~\cite{IVANOV1987612,PhysRevLett.93.016403}, and their 
potential to condense into a degenerate state, remain open questions.

\begin{figure}[tbp]
\includegraphics[width=0.48\textwidth]{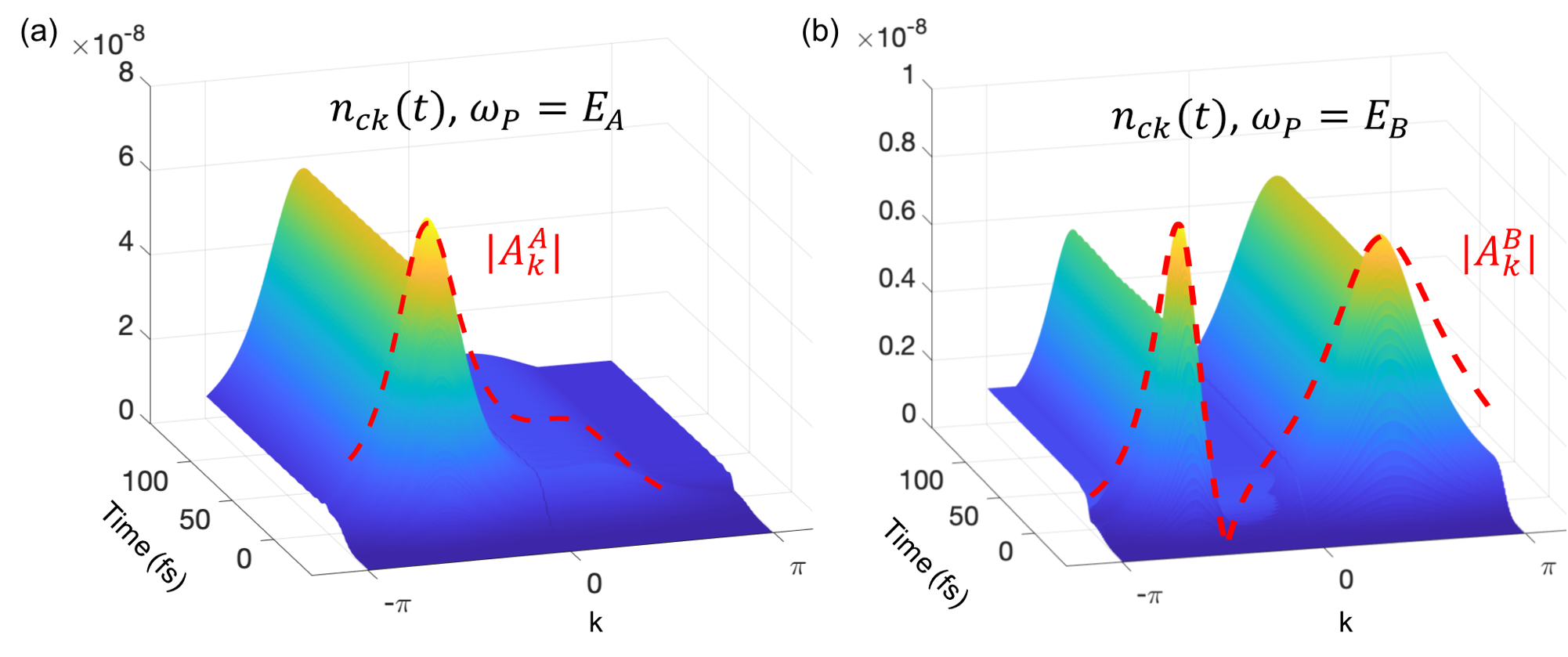}
\caption{Momentum resolved carrier occupations $n_k^c(t)$ versus time 
for resonant excitations with (a) the A exciton 
[$\w_{P}=E_{A}=0.54$~eV] and 
(b) the B exciton [$\w_{P}=E_{B}=1.03$~eV].}   
\label{fig:nck} 
\end{figure}

In Fig.~\ref{fig:nck}(a) we display the momentum-resolved 
distribution of carriers in the conduction band, see 
Eq.~(\ref{nckanalyt}). For negative times $n_{ck}(t)$ grows 
proportional to the modulus square of the A-exciton wavefunction 
$|A^{A}_{k}|^{2}$. For $t>0$, the distribution  slightly spreads 
due to phonon dressing.     

If the probe pulse duration $T_p$ is much longer than typical 
electronic timescales, and its frequency is sufficiently large to 
resolve the desired removal energies, the excitonic impact on
TR-ARPES in both coherent and incoherent regimes can be 
numerically calculated from [cfr. Eq.~(\ref{spectralf})]       
\begin{equation}
A_k(T, \omega)=-i \int_{T-\frac{\tau_p}{2}}^{T+\frac{\tau_p}{2}} d 
\tau e^{i \omega \tau} G_{cck}\left(T+\frac{\tau}{2}, 
T-\frac{\tau}{2}\right),  
\label{spectrnum}
\end{equation}
where 
$T$ is the time at which the probe interacts with the system
and $\omega$ is the shifted photoemission energy. 
Henceforth, we refer to $T$ as the pump-probe delay.
% Details on the numerical implementation of Eq.~(\ref{spectrnum}) are 
% given in Appendix~CCC. 

\begin{figure*}[htb]
\includegraphics[width=1.03\textwidth]{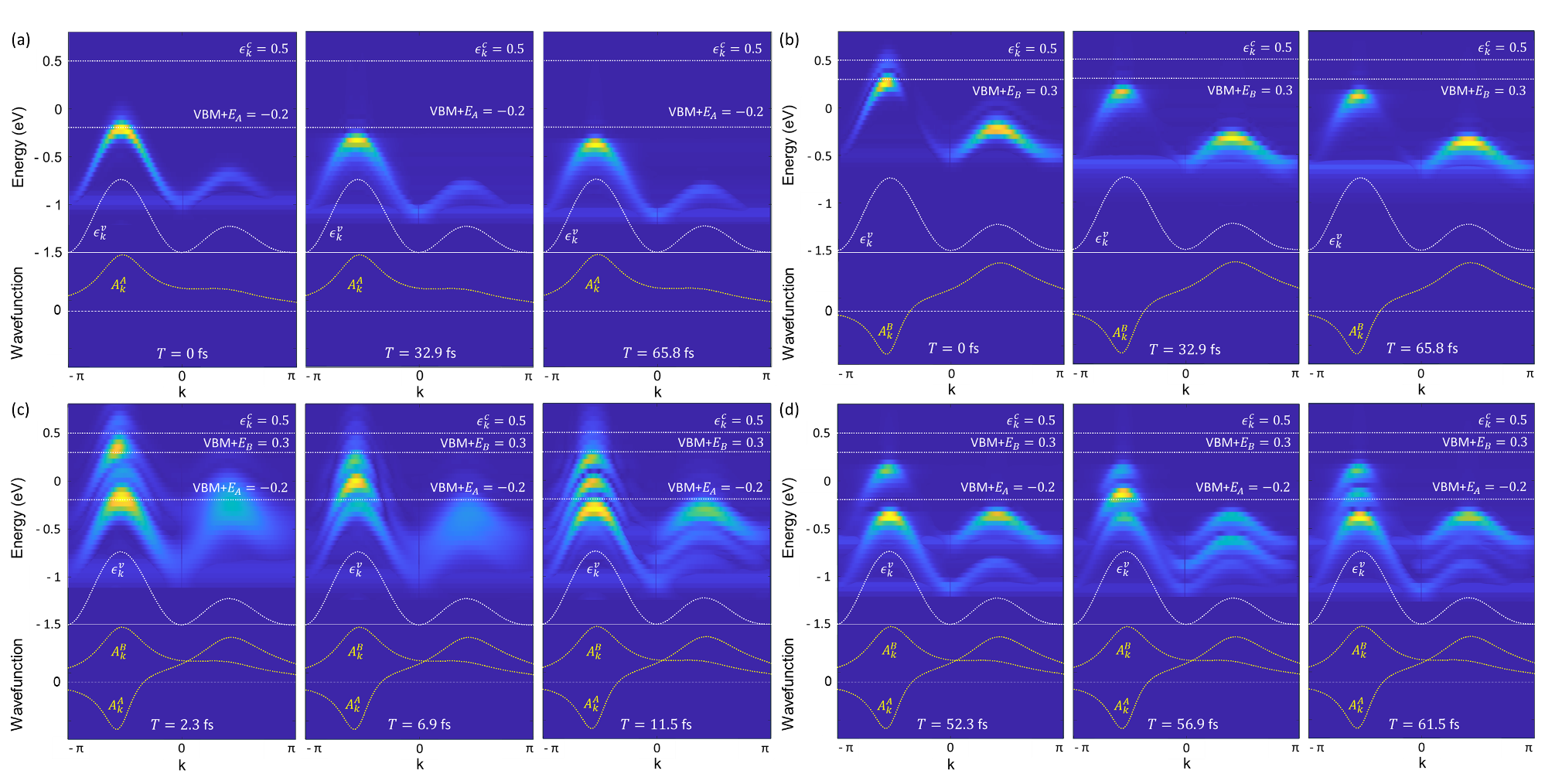}
\caption{TR-ARPES spectra for different pump-probe delays $T$. 
The pump is resonant either with the A 
exciton [$\w_{P}=E_{A}=0.54$~eV] (a) or the B exciton 
[$\w_{P}=E_{B}=1.03$~eV] (b). (c-d) Nonresonant spectra
[$\w_{P}=0.99$~eV]. In each panel the electronic bands (top, dashed 
lines) and 
exciton wavefunctions (bottom) are displayed.} 
\label{fig:A}
\end{figure*}

We choose the probe duration $T_p=80$~fs, significantly 
longer than the excitonic timescales $2\pi/E_\l$ for both the A- and 
B-exciton. The TR-ARPES spectra is shown in 
Fig.~\ref{fig:A}(a). The exact solution allows us to follow the 
temporal evolution of the spectral properties. In particular, in the 
overlapping regime (zero pump-probe delay) we clearly distinguish the 
excitonic replica of the valence band shifted upward by 
the A-exciton energy $E_{A}$~\cite{perfetto_first-principles_2016}.
The spectral weight of the replica is 
proportional to the modulus square of the excitonic 
wavefunctions$|A^{A}_{k}|^{2}$, 
and its position is slightly renormalized by the optical field. As the 
system enters the incoherent regime ($T\simeq  66$~fs), the excitonic 
replica exhibits a downward Stokes 
shift~\cite{toyozawa2003optical,liu_exciton-polaron_2021,dai_theory_2024} 
and broadens.
This is consistent with the fact that incoherent excitons are 
excitons dressed by phonons (or exciton-polarons), see Eq.~(\ref{incex}).

Pumping in resonance with the B-exciton 
yields a similar behavior. In fact, phonon-mediated scattering 
between different exciton species is not allowed in the investigated model. 
As already observed, the exciton-phonon 
coupling for the transition $BQ\to AQ'$ vanishes, see 
Eq.~(\ref{exphoncoup}).

We have performed simulations with pump frequency $\w_{P}=E_{B}$ and pump 
duration $T_{P}\simeq 40$~fs.
In Fig.~\ref{fig:nck}(b) we display the evolution of the momentum resolved 
carrier density, highlighting the profile of the modulus square 
of the B-exciton 
wavefunction.  In Fig.~\ref{fig:A}(b) we display the evolution of the 
TR-ARPES spectrum. We observe the excitonic replica at energy $E_{B}$ 
above the VBM, and the Stokes shift in the incoherent regime.
Since $|A^{B}_{k}|^{2}$ has two well 
pronounced maxima in correspondence of the local valleys of the 
valence band, the signal intensities from both valleys are comparable.

\begin{figure}[tbp]
\includegraphics[width=0.48\textwidth]{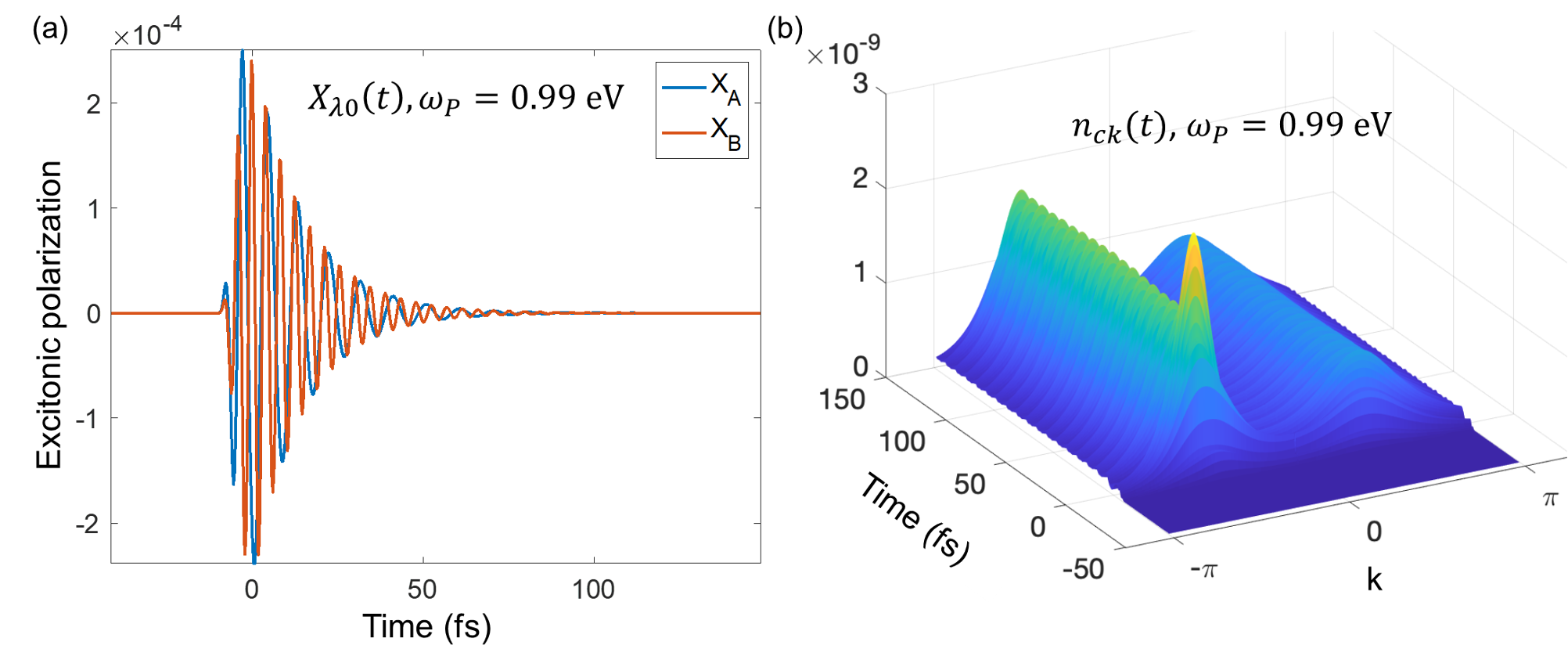}
\caption{(a) Exciton polarizations and (b) momentum-resolved carrier 
occupation versus time for nonresonant pumping.}       
\label{fig:nonres}
\end{figure}

\subsection{Nonresonant pumping}

We now address the nonresonant case. We tune the pump frequency 
between the A- and B-exciton energy, $\omega_P=0.99$~eV, and set the 
pump duration $T_p=10$~fs. 
In Fig.~\ref{fig:nonres}(a) we show the excitonic 
polarizations of the two excitons. The decay rates are identical
due to the diagonal exciton-phonon coupling, 
see Eq.~(\ref{exphoncoup}), and because the electron-phonon 
interaction depends solely on the momentum transfer. Interestingly, 
for the model under investigation, the decay rates predicted by the 
excitonic Bloch equations are also identical. 

The momentum-resolved carrier distribution exhibits  
oscillations surviving in the incoherent regimes, see  
Fig.~\ref{fig:nonres}(b). Although the G-X and X-X 
coherences vanish after $\sim 66$~fs, the X-X coherence transforms 
into a X$^{\rm inc}$-X$^{\rm inc}$ coherence. 
In the incoherent 
regime carriers bounce back and forth between the two incoherent 
exciton states, giving rise to persistent oscillations.

Figure~\ref{fig:A}(c,d) illustrates the time-resolved spectral function 
in the overlapping  ($T<40$~fs) and nonoverlapping ($T>40$~fs) regimes. 
As pointed out in Ref.~\cite{perfetto_time-resolved_2020}, capturing 
the G-X coherence in TR-ARPES necessitates the use of ultrashort probe pulses. 
Therefore, we here focus on the X-X coherence.
In the absence of electron-phonon coupling, this situation has been 
analyzed in Ref.~\cite{rustagi_coherent_2019}: one would observe  
three excitonic replica, two located at energies $E_A$ and 
$E_{B}$ above the VBM, with intensities proportional to the square 
modulus of the excitonic wavefunctions, and a third 
at energy
$(E_A+E_B)/2$ above the VBM, with an intensity that oscillates with period  
$2\pi/(E_B-E_A)$. The quantum beats of the third excitonic replica are 
the signature of X-X coherence in TR-ARPES. In the presence of 
electron-phonon coupling our results show that the X-X coherence gradually 
transforms into a robust X$^{\rm inc}$-X$^{\rm inc}$ coherence, which 
persists indefinitely, indicating that
the system does not attain a steady state.            
The conversion X-X$\to$X$^{\rm inc}$-X$^{\rm 
inc}$ is supported by several lines of evidence: (i) no coherent 
excitons exist for $t>66$~fs, see Fig.~\ref{fig:nonres}(a) and (ii) 
the position of the three excitonic replicas moves downward in energy 
due to the Stokes shift.

The different outcome of all observable quantities between 
the resonant and nonresonant pumping scenarios is 
rooted in the many-body reduced  density matrix, see Eq.~(\ref{eldm}). In the 
long-time limit the G-X and X-X coherences vanish, and therefore
\begin{align}
\lim_{t \rightarrow \infty}&\hat{\rho}_{\mathrm{el}}(t)= 
|\alpha(0)|^2|\Phi_g\ket\bra\Phi_g| 
+\sum_{\lambda\lambda'\blQ}N_{\lambda\lambda'\blQ}^{\text{inc}}(t)
|X_{\lambda'\blQ}\ket\bra X_{\lambda\blQ}|.
\label{asymrhoel}
\end{align}
The long-time limit behavior of the incoherent excitonic 
density matrix is, see Eq.~(\ref{Nllp}), 
\begin{align}
N_{\lambda\lambda'\blQ}(t\to 
\iif)=N_{\lambda\lambda'\blQ}^{\iif}e^{-i(E_{\l}-E_{\l'})t},
\end{align}
where the asymptotic (time-independent) matrix $N_{\lambda\lambda'\blQ}^{\iif}$ is 
history-dependent. 
Pumping resonantly with the $\l_{0}$ exciton we find  
$N_{\lambda\lambda'\blQ}^{\iif}=\d_{\l\l_{0}}\d_{\l'\l_{0}}N_{\l_{0}\blQ}$, 
and therefore $\hat{\rho}_{\mathrm{el}}$ in Eq.~(\ref{asymrhoel}) as well as all 
observable quantities
attain a constant value.  
In contrast, for nonresonant pumping $N_{\lambda\lambda'\blQ}^{\iif}$ 
is an off-diagonal matrix. 
These results suggest that the 
X$^{\rm inc}$-X$^{\rm 
inc}$ coherence follows different dynamics than those governing the 
G-X and X-X coherences. 

\section{Summary and Conclusions}
\label{concsec}

We have considered a two-band semiconductor model Hamiltonian 
that includes both electron-electron and electron-phonon interactions. 
Under simplifying assumptions, such as a flat conduction band and the 
neglect of electron-phonon scattering in the valence band, 
we have derived several exact results in the weak pumping 
scenario. Specifically, we  
present analytic expressions for the many-body state, the 
one-particle Green's function, and the exciton Green's function.

The explicit form of the many-body state provides a rigorous 
mathematical definition of concepts such as coherent and incoherent excitons.  
In particular, coherent excitons are quantum superpositions of  
ground state and pure exciton states, whereas incoherent excitons 
are  exciton states that are dressed by a phonon cloud 
(exciton-polaron). 

Below gap optical excitations generate coherent excitons, and hence 
G-X and X-X coherence. Phonon-induced dephasing suppresses 
G-X coherence and converts X-X coherence into 
a coherence between incoherent excitons. For the investigated model 
the X$^{\rm inc}$-X$^{\rm inc}$ coherence is resistant to 
phonon-induced dephasing and persists indefinitely. 
Our simulations show that this coherence generate quantum beats in 
the so called incoherent regime (zero excitonic polarization) in both 
carrier populations and TR-ARPES signal.

Although the investigated model is a simplification of the 
first-principles Hamiltonian of real semiconductors, it 
correctly captures key universal features, 
such as the the formation of coherent excitons and their subsequent 
decay into incoherent (bright and dark) excitons.
The infinite lifetime of the X$^{\rm inc}$-X$^{\rm inc}$ coherence
is obvioulsy an artifact of the simplifications inherent in the model 
Hamiltonian. Specifically, it results from the fact that the exciton 
wave functions are independent of the center-of-mass momentum, 
and the electron-phonon coupling depends only on the momentum 
transfer. Furthermore, we are here discarding
intervalley scatterings as well as relaxation toward lower energy dark 
excitons. Nonetheless, for systems with lowest energy excitons
localized in one valley, the inclusion of the aforementioned 
effects  are only weakly detrimental.
Our results highlight that the X$^{\rm inc}$–X$^{\rm inc}$ coherence 
evolves on a different timescale than the excitonic coherences 
studied to date, and it may be significantly longer.    
The long-lived (several picoseconds) coherence observed in time-resolved optical 
experiments~\cite{bar-ad_absorption_1991,sim_ultrafast_2018,trifonov_exciton_2020} 
may be explained in this context.
Learning how to control and manipulate 
excitonic coherence could also enable the exploration of 
exciton-driven Floquet matter for quantum 
technology.

\acknowledgments

We acknowledge funding from Ministero Universit\`a e 
Ricerca PRIN under grant agreement No. 2022WZ8LME, 
from INFN through project TIME2QUEST, 
from European Union’s Horizon
Europe research and innovation programme under the Marie
Sk{\polishl}odowska-Curie Doctoral Networks grant agreement No. 101118915 –
TIMES, 
and from Tor Vergata University through project TESLA.

\appendix

\begin{widetext}

\section{Analytic Time-Dependent Many-Body State}
\label{aapp}

Starting from the termination of the external pump field, the many-body state evolves like
\begin{align}
|\Psi(t>0)\rangle
=&\alpha(0)|\Phi_g\rangle
+\sum_\l
\b_\l(0)e^{-i\hat{H}t}
|X_{\l\bz}\rangle.
\end{align}
We expand the evolved state by a total momentum conserved phonon dressed exciton states
\begin{align}
e^{-i\hat{H}t}
|X_{\l\bz}\rangle
=&\sum_{M=0}^\infty
\frac{1}{M!}
\sum_{\l\blq_1\ldots\blq_M}
\hat{b}^{\dag}_{\blq_1}\ldots
\hat{b}^{\dag}_{\blq_M}
|X_{\l-\blq_1\ldots-\blq_M}\rangle
\langle X_{\l-\blq_1\ldots-\blq_M}|
\hat{b}_{\blq_M}\ldots
\hat{b}_{\blq_1}
e^{-i\hat{H}t}
|X_{\l\bz}\rangle,
\label{tdstate}
\end{align}
with orthonormality relations
\begin{align}
\bra X_{\l\blQ}|X_{\l'\blQ'}\ket
&=\bra \Q_g|\hat{X}_{\l\blQ}\hat{X}^{\dag}_{\l'\blQ'}|\Q_g\ket\nn\\
&=\sum_{\blk\blk'}
A^{\l*}_{\blk}
A^{\l'}_{\blk'}
\bra \Q_g|\hat{v}^{\dag}_{\blk}\hat{c}_{\blk+\blQ}
\hat{c}^{\dag}_{\blk'+\blQ'}\hat{v}_{\blk'}|\Q_g\ket\nn\\
&=\d_{\l\l'}\d_{\blQ,\blQ'},
\end{align}
and
\begin{align}
\bra X_{\l\blQ}|
\hat{b}_{\blq_M}\ldots
\hat{b}_{\blq_1}
\hat{b}^{\dag}_{\blq_1'}\ldots
\hat{b}^{\dag}_{\blq_{M'}'}
|X_{\l'\blQ'}\ket=
\d_{MM'}\d_{\l\l'}\d_{\blQ,\blQ'}
\sum_P\prod_{j=1}^M \d_{\blq_j,\blq_{P(j)}'}.
\label{orth_nor_rel_Xph}
\end{align}
From Eq.~(\ref{tdstate}) 
we can generate a hierarchy of differential equations for the amplitudes
\begin{align}
&\quad i\frac{d}{dt}
\langle X_{\l-\blq_1\ldots-\blq_M}|
\hat{b}_{\blq_M}\ldots
\hat{b}_{\blq_1}
e^{-i\hat{H}t}
|X_{\l\bz}\rangle\nn\\
&=\langle X_{\l-\blq_1\ldots-\blq_M}|
\hat{b}_{\blq_M}\ldots
\hat{b}_{\blq_1}\hat{H}
e^{-i\hat{H}t}
|X_{\l\bz}\rangle\nn\\
&=(E_\l+\sum_{j=1}^M\w_{\blq_j})\langle X_{\l-\blq_1\ldots-\blq_M}|
\hat{b}_{\blq_M}\ldots
\hat{b}_{\blq_1}
e^{-i\hat{H}t}
|X_{\l\bz}\rangle\nn\\
&\quad+\sum_{j=1}^M
\frac{g_{\blq_j}}{\sqrt{\callN}}
\langle X_{\l-\blq_1\ldots -\stackrel{\sqcap}{\blq}_j \ldots-\blq_M}|
\hat{b}_{\blq_M}\ldots
\stackrel{\sqcap}{\hat{b}}_{\blq_j}\ldots
\hat{b}_{\blq_1}
e^{-i\hat{H}t}
|X_{\l\bz}\rangle\nn\\
&\quad+\sum_{\blq}
\frac{g_{\blq}}{\sqrt{\callN}}
\langle X_{\l-\blq-\blq_1\ldots -\blq_M}|
\hat{b}_{\blq_M}\ldots
\hat{b}_{\blq_1}
\hat{b}_{\blq}
e^{-i\hat{H}t}
|X_{\l\bz}\rangle.
\end{align}
If we look for solutions of the form
\begin{align}
\langle X_{\l-\blq_1\ldots-\blq_M}|
\hat{b}_{\blq_M}\ldots
\hat{b}_{\blq_1}
e^{-i\hat{H}t}
|X_{\l\bz}\rangle
=f_{\blq_1}(t)\ldots f_{\blq_M}(t)
\ell(t)e^{-iE_\l t},
\end{align}
the dynamics of excitons and phonons is decoupled 
and the hierarchy is solved analytically by solving
\begin{equation}
\begin{aligned}
i\dot{f}_\blq(t)&=\frac{g_\blq}{\sqrt{\callN}}+\w_\blq f_\blq(t),\\
i\dot{\ell}(t)&=\sum_\blq \frac{g_\blq}{\sqrt{\callN}} f_\blq(t)\ell(t).
\label{f_ell}
\end{aligned}
\end{equation}
The solutions of Eq.~(\ref{f_ell}) with boundary condition $f_\blq(0)=0$ and $\ell(0)=1$ are
\begin{equation}
\begin{aligned}
f_\blq(t)&=\frac{1}{\sqrt{\mathcal{N}}} 
\frac{g_\blq}{\omega_\blq}\left(e^{-i \omega_\blq t}-1\right),\\
\ell(t)&=\exp \left[\frac{1}{\mathcal{N}} 
\sum_\blq\left(\frac{g_\blq}{\omega_\blq}\right)^2\left(-1+e^{-i 
\omega_\blq t}+i \omega_\blq t\right)\right].
\end{aligned}
\end{equation}
Therefore, 
the analytic expression of the time-dependent many-body state of the 
system for positive time is
\begin{align}
|\Q(t>0)\ket=\alpha(0)|\F_g\ket+
\sum_\l\b_\l(0)\ell(t)e^{-iE_\l t}
\sum_{M=0}^\infty\frac{1}{M!}
\sum_{\blq_1\ldots\blq_M}
f_{\blq_1}(t)\ldots f_{\blq_M}(t)
\hat{b}^{\dag}_{\blq_1}\ldots
\hat{b}^{\dag}_{\blq_M}
|X_{\l-\blq_1\ldots-\blq_M}\ket.
\label{App_exact_wf1}
\end{align}

We separate out the contribution with zero phonons from the second 
term    
\begin{align}
|\Q(t>0)\ket=\alpha(0)|\F_g\ket+
\sum_\l\b_\l(0)\ell(t)e^{-iE_\l t}
\left(|X_{\l\bz}\ket+\sum_\blQ|\bar{X}_{\l\blQ}^\text{inc}(t)\ket\right),
\label{App_exact_wf2}
\end{align}
where the definition of coherent exciton states $|X_{\l\bz}\ket$ is 
given in Eq.~(\ref{fqex}), and the unnormalized incoherent exciton states are given by
\begin{align}
|\bar{X}_{\l \blQ}^{\text{inc}}(t)\rangle
=\sum_{M=1}^\infty
\frac{1}{M!}
\sum_{ \blq_1 \ldots \blq_M}
\delta_{\blq_1\ldots +\blq_M,-\blQ}
f_{\blq_1}(t) \ldots f_{\blq_M}(t)
\hat{b}^{\dag}_{\blq_1}\ldots\hat{b}^{\dag}_{\blq_M}
|{X}_{\l\blQ}\rangle.
\label{Xbar}
\end{align}
According to Eq.~(\ref{orth_nor_rel_Xph}) the coherent and incoherent exciton states are  orthogonal
\begin{align}
\bra X_{\l\bz}|\bar{X}^{\rm inc}_{\l'\blQ'}\ket=0.
\end{align}
Let us normalize the incoherent exciton states. The inner product  is
\begin{align}
\bra \bar{X}_{\l \blQ}^{\text{inc}}(t)|\bar{X}_{\l' 
\blQ'}^{\text{inc}}(t)\ket
&=\sum_{M,M'=1}^\infty\frac{1}{M!M'!}
\sum_{\substack{\blq_1\ldots\blq_M\\\blq_1'\ldots\blq_{M'}'}}
\d_{\blq_1\ldots +\blq_M,-\blQ}
\d_{\blq_1'\ldots +\blq_{M'}',-\blQ'}
f_{\blq_1'}(t) \ldots f_{\blq_{M'}'}(t)
f^*_{\blq_1}(t) \ldots f^*_{\blq_{M}}(t)
\nn\\
&\quad\times \bra X_{\l\blQ}|
\hat{b}_{\blq_M}\ldots
\hat{b}_{\blq_1}
\hat{b}^{\dag}_{\blq_1'}\ldots
\hat{b}^{\dag}_{\blq_{M'}'}
|X_{\l'\blQ'}\ket.
\end{align}
Taking into account Eq.~(\ref{orth_nor_rel_Xph}), we have
\begin{align}
\bra \bar{X}_{\l \blQ}^{\text{inc}}(t)|\bar{X}_{\l' 
\blQ'}^{\text{inc}}(t)\ket
&=\d_{\l\l'}\d_{\blQ,\blQ'}\sum_{M=1}^\infty\frac{1}{M!}
\sum_{{\blq_1\ldots\blq_M}}
\d_{\blq_1\ldots +\blq_M,-\blQ}
|f_{\blq_1}(t)|^2 \ldots |f_{\blq_M}(t)|^2.
\end{align}
We rewrite the Kronecker delta function in terms of its discrete 
Fourier transform 
\begin{align}
\d_{\blq_1\ldots +\blq_M,-\blQ}=\frac{1}{\callN}\sum_{\bln}e^{-i(\blq_1\ldots +\blq_M-\blQ)\cdot\bln},
\end{align}
where $\bln$ are unit cell vectors. 
The inner product then takes the compact form
\begin{align}
\bra \bar{X}_{\l \blQ}^{\text{inc}}(t)|\bar{X}_{\l' 
\blQ'}^{\text{inc}}(t)\ket
&=\d_{\l\l'}\d_{\blQ,\blQ'}
\frac{1}{\callN}\sum_{\bln}e^{i\blQ\cdot\bln}
\sum_{M=1}^\infty\frac{1}{M!}
\left(\sum_{{\blq}}
|f_{\blq}(t)|^2 e^{-i\blq\cdot \bln}\right)^M=\d_{\l\l'}\d_{\blQ,\blQ'}S_\blQ(t),
\end{align}
where
\begin{align}
S_\blQ(t)=
\frac{1}{\mathcal{N}}
\sum_\bln e^{i\blQ\cdot\bln}
\exp{\left[
\sum_\blq 
|f_\blq(t)|^{2}
e^{-i\blq\cdot\bln}
\right]} 
-\delta_{\blQ,\bz}.
\end{align}
Hence the function $S_\blQ(t)$ is real and nonnegative for all times. 
Normalizing the incoherent exciton state 
as      
\begin{align}
|{X}_{\l \blQ}^{\text{inc}}(t)\ket=
\frac{1}{\sqrt{S_\blQ(t)}}
|\bar{X}_{\l \blQ}^{\text{inc}}(t)\ket,
\end{align}
we eventually  obtain the time-dependent many-body state
\begin{align}
|\Psi(t>0)\rangle
&=\alpha(0)|\Phi_g\rangle
+\sum_\l
\b_\l(0)\ell(t)e^{-iE_\l t}\left(
|X_{\l\bz}\rangle+\sum_{\blQ}\sqrt{S_{\blQ}(t)}|X_{\l \blQ}^{\text{inc}}(t)\rangle\right)
\end{align}
with orthonormal relations
\begin{subequations}
\begin{align}
\bra X_{\l\bz}|X_{\l'\bz}\ket&=\d_{\l\l'},
\\
\bra X_{\l \blQ}^{\text{inc}}(t)|X_{\l' 
\blQ'}^{\text{inc}}(t)\ket&=\d_{\l\l'}\d_{\blQ,\blQ'},
\\
\bra X_{\l\bz}|X_{\l' 
\blQ'}^{\text{inc}}(t)\ket&=0.
\end{align}
\end{subequations}

\section{Green's Function}
\label{bapp}

\subsection{One-particle Green's Function}
We calculate the one-particle lesser Green's function defined in 
Eqs.~(\ref{GF1}) and (\ref{GF2}). For the coherent 
contribution   we have
\begin{align}
G_{cc\blk}^{\text{coh}}
(t, t^{\prime})=&i\bra \Psi(t^{\prime})|\hat{P}^{\rm coh}
\hat{c}_\blk^{\dagger} 
e^{-i \hat{H}(  t^{\prime}-t)}
\hat{c}_\blk\hat{P}^{\rm coh}
|\Psi(t)\rangle 
\nn\\
=&i\sum_{\l\l'}\b_\l(t)\b_{\l'}^*(t')
\ell(t)\ell^*(t')e^{-iE_\l t}e^{iE_{\l'}t'}
\bra X_{\l'\bz}|\hat{c}_\blk^{\dagger} 
e^{-i \hat{H}(  t^{\prime}-t)}
\hat{c}_\blk |X_{\l\bz}\ket.
\end{align}
The state $\hat{c}_\blk |X_{\l\bz}\ket$ is an eigenstate 
of  $\hat{H}$ with a $\blk$-momentum hole:
\begin{align}
e^{-i \hat{H}(  t^{\prime}-t)}
\hat{c}_\blk |X_{\l\bz}\ket
&=A_\blk^\l e^{-i(-\e_\blk^v)(t'-t)}\hat{v}_\blk |\F_g\ket.
\end{align}
Hence the coherent contribution of the one-particle Green's function is given by
\begin{align}
G_{cc\blk}^{\text{coh}}
(t, t^{\prime})
=&i
\sum_{\l\l'}\b_{\l}(t)\b^{\ast}_{\l'}(t')
\ell(t)\ell^{\ast}(t')
A_\blk^{\l} A^{\l'\ast}_\blk
e^{-i(\epsilon_\blk^v+E_\l) t}
e^{i(\epsilon_\blk^v+E_{\l'})t'}.
\end{align}

For the incoherent contribution we have
\begin{align}
G_{cc\blk}^{\rm inc}
(t, t^{\prime})=&i\bra \Q(t^{\prime})|\hat{P}^{\rm inc}
\hat{c}_\blk^{\dagger} 
e^{-i \hat{H}(  t^{\prime}-t)}
\hat{c}_\blk\hat{P}^{\rm inc}
|\Q(t)\ket 
\nn\\
=&i\sum_{\l\l'}\b_\l(t)\b_{\l'}^*(t')
\ell(t)\ell^*(t')e^{-iE_\l t}e^{iE_{\l'}t'}
\sum_{\blQ\blQ'}
\bra \bar{X}^{\rm inc}_{\l'\blQ'}|\hat{c}_\blk^{\dagger} 
e^{-i \hat{H}(  t^{\prime}-t)}
\hat{c}_\blk |\bar{X}^{\rm inc}_{\l\blQ}\ket.
\label{App_GF_inc}
\end{align}
Using Eq.~(\ref{Xbar}), the inner product reads
\begin{align}
\bra \bar{X}_{\l' \blQ'}^{\text{inc}}(t')|
\hat{c}_\blk^{\dagger} 
e^{-i \hat{H}(  t^{\prime}-t)}
\hat{c}_\blk 
|\bar{X}_{\l 
\blQ}^{\text{inc}}(t)\ket
&=\sum_{M,M'=1}^\infty\frac{1}{M!M'!}
\sum_{\substack{\blq_1\ldots\blq_M\\\blq_1'\ldots\blq_{M'}'}}
\d_{\blq_1\ldots +\blq_M,-\blQ}
\d_{\blq_1'\ldots +\blq_{M'}',-\blQ'}
f^*_{\blq_1'}(t') \ldots f^*_{\blq_{M'}'}(t')
f_{\blq_1}(t) \ldots f_{\blq_M}(t)
\nn\\
&\quad\times \bra X_{\l'\blQ'}|
\hat{b}_{\blq_{M'}'}\ldots
\hat{b}_{\blq_1'}
\hat{c}_\blk^{\dagger} 
e^{-i \hat{H}(  t^{\prime}-t)}
\hat{c}_\blk 
\hat{b}^{\dag}_{\blq_1}\ldots
\hat{b}^{\dag}_{\blq_{M}}
|X_{\l\blQ}\ket.
\end{align}
The phonon dressed hole state is also an eigenstate of $\hat{H}$:
\begin{align}
e^{-i \hat{H}(  t^{\prime}-t)}
\hat{c}_\blk 
\hat{b}^{\dag}_{\blq_1}\ldots
\hat{b}^{\dag}_{\blq_{M}}
|X_{\l\blQ}\ket
&=A^\l_{\blk-\blQ}
e^{-i(-\e^v_{\blk-\blQ} +\w_{\blq_1}\ldots +\w_{\blq_M})(t'-t)}
\hat{b}^{\dag}_{\blq_1}\ldots
\hat{b}^{\dag}_{\blq_{M}}
\hat{v}_{\blk-\blQ}|\F_g\ket.
\end{align}
Then, we have
\begin{align}
\bra \bar{X}_{\l' \blQ'}^{\text{inc}}(t')|
\hat{c}_\blk^{\dagger} 
e^{-i \hat{H}(  t^{\prime}-t)}
\hat{c}_\blk 
|\bar{X}_{\l 
\blQ}^{\text{inc}}(t)\ket
&=\sum_{M,M'=1}^\infty\frac{1}{M!M'!}
\sum_{\substack{\blq_1\ldots\blq_M\\\blq_1'\ldots\blq_{M'}'}}
\d_{\blq_1\ldots +\blq_M,-\blQ}
\d_{\blq_1'\ldots +\blq_{M'}',-\blQ'}
f^*_{\blq_1'}(t') \ldots f^*_{\blq_{M'}'}(t')
f_{\blq_1}(t) \ldots f_{\blq_M}(t)
\nn\\
&\quad\times A^{\l'*}_{\blk-\blQ'}
A^\l_{\blk-\blQ}
e^{-i(-\e^v_{\blk-\blQ} +\w_{\blq_1}\ldots +\w_{\blq_M})(t'-t)}
 \bra \F_g|
\hat{v}^{\dag}_{\blk-\blQ'}
\hat{b}_{\blq_{M'}'}\ldots
\hat{b}_{\blq_1'}
\hat{b}^{\dag}_{\blq_1}\ldots
\hat{b}^{\dag}_{\blq_{M}}
\hat{v}_{\blk-\blQ}
|\F_g\ket.
\label{GF_S_1}
\end{align}
We use again the relation in Eq.~(\ref{orth_nor_rel_Xph}) to obtain
\begin{align}
\bra \F_g|
\hat{v}^{\dag}_{\blk-\blQ'}
\hat{b}_{\blq_{M'}'}\ldots
\hat{b}_{\blq_1'}
\hat{b}^{\dag}_{\blq_1}\ldots
\hat{b}^{\dag}_{\blq_{M}}
\hat{v}_{\blk-\blQ}
|\F_g\ket
&=\d_{MM'}\d_{\blQ,\blQ'}
\sum_P\prod_{j=1}^M\d_{\blq_j,\blq_{P(j)}'}.
\end{align}
Summing over all momenta $\blq_1,\ldots,\blq_M$, and reorganizing the 
factors $f_\blq$ and $e^{-i\w_q(t'-t)}$, we have
\begin{align}
\bra \bar{X}_{\l' \blQ'}^{\text{inc}}(t')|
\hat{c}_\blk^{\dagger} 
e^{-i \hat{H}(  t^{\prime}-t)}
\hat{c}_\blk 
|\bar{X}_{\l 
\blQ}^{\text{inc}}(t)\ket
&=
\d_{\blQ,\blQ'}
\sum_{M=1}^\infty\frac{1}{M!}
\sum_{{\blq_1\ldots\blq_M}}
\d_{\blq_1\ldots +\blq_M,-\blQ}
\left[\prod_{j=1}^M
f^*_{\blq_j}(t')f_{\blq_j}(t)e^{-i\w_{\blq_j}(t'-t)}\right] 
A^{\l'*}_{\blk-\blQ}
A^\l_{\blk-\blQ}
e^{-i(-\e^v_{\blk-\blQ} )(t'-t)},
\label{GF_S_2}
\end{align}
from which
\begin{align}
\sum_{M=1}^\infty\frac{1}{M!}
\sum_{{\blq_1\ldots\blq_M}}
\d_{\blq_1\ldots +\blq_M,-\blQ}
\left[\prod_{j=1}^M
f^*_{\blq_j}(t')f_{\blq_j}(t)
e^{-i\w_{\blq_j}(t'-t)}\right]
=&\frac{1}{\callN}\sum_\bln
e^{i\blQ\cdot\bln}
\sum_{M=1}^\infty\frac{1}{M!}
\left(
\sum_{{\blq}}
f^*_{\blq}(t')f_{\blq}(t)e^{-i\w_{\blq}(t'-t)-i\blq\cdot\bln}
\right)^M\nn\\
=&\frac{1}{\mathcal{N}}
\sum_\bln e^{i\blQ \cdot\bln}
\exp{\left[
\sum_\blq 
f^*_\blq(t') f_\blq(t)
e^{-i\omega_\blq(t'-t)-i\blq\cdot\bln}
\right]} -\delta_{\blQ,\bz}\nn\\
\equiv &S_\blQ(t,t').
\label{GF_S_3}
\end{align}
In conclusion,
\begin{align}
\bra \bar{X}_{\l' \blQ'}^{\text{inc}}(t')|
\hat{c}_\blk^{\dagger} 
e^{-i \hat{H}(  t^{\prime}-t)}
\hat{c}_\blk 
|\bar{X}_{\l 
\blQ}^{\text{inc}}(t)\ket
&=
A^{\l'*}_{\blk-\blQ}
A^\l_{\blk-\blQ}
e^{-i(-\e^v_{\blk-\blQ} )(t'-t)}
\d_{\blQ,\blQ'}S_\blQ(t,t'),
\label{GF_S_4}
\end{align}
and substitution into Eq.~(\ref{App_GF_inc}) yields the sought expression
\begin{align}
G_{cc\blk}^{\rm inc}
(t, t^{\prime})
=&\sum_\blQ S_\blQ(t,t')
\left[i\sum_{\l\l'}\b_\l(t)\b_{\l'}^*(t')
\ell(t)\ell^*(t')
A^{\l'*}_{\blk-\blQ}
A^\l_{\blk-\blQ}
e^{-i(E_\l+\e^v_{\blk-\blQ})t}
e^{i(E_{\l'}+\e^v_{\blk-\blQ})t'}
\right]\nn\\
=&\sum_\blQ S_\blQ(t,t')G_{cc\blk-\blQ}^{\rm coh}
(t, t^{\prime}).
\end{align}

\subsection{Excitonic Green's Function}

We start from the coherent contribution in Eq.~(\ref{X_GF1})
\begin{align}
N^{\text{coh}}_{\lambda\lambda'}
(t, t^{\prime})
&=\bra\Q(t')|\hat{P}^{\rm coh} 
\hat{X}_{\lambda' \bz}^{\dagger} 
e^{-i \hat{H}( t^{\prime}-t)} 
\hat{X}_{\lambda \bz}
\hat{P}^{\rm coh}|\Q(t)\ket\nn\\
=&\sum_{\l_1\l_2}\b_{\l_1}(t)\b_{\l_2}^*(t')
\ell(t)\ell^*(t')e^{-iE_{\l_1} t}e^{iE_{{\l_2}}t'}
\bra X_{\l_2\bz}|
\hat{X}_{\lambda' \bz}^{\dagger} 
e^{-i \hat{H}( t^{\prime}-t)} 
\hat{X}_{\lambda \bz}
|X_{\l_1\bz}\ket.
\end{align}
From the anti-commutation relation of electrons and orthonormalization of excitonic wavefunctions we have
\begin{align}
\hat{X}_{\l\blQ}|X_{\l'\blQ'}\ket
&=\sum_{\blk\blk'}A^{\l *}_\blk
A^{\l'}_{\blk'}\hat{v}_{\blk}^{\dag}\hat{c}_{\blk+\blQ}
\hat{c}_{\blk'+\blQ'}^{\dag}\hat{v}_{\blk'}|\F_g\ket\nn\\
&=\d_{\l\l'}\d_{\blQ,\blQ'}|\F_g\ket.
\end{align}
Hence the coherent contribution is
\begin{align}
N^{\text{coh}}_{\lambda\lambda'}
(t, t^{\prime})
=&\b_\l(t)\b_{\l'}^*(t')
\ell(t)\ell^*(t')e^{-iE_{\l} t+iE_{{\l'}}t'}.
\end{align}
Considering Eq.~(\ref{xlcoh}) and $|\a(0)|^2\simeq 1$ in linear regime, we have
\begin{align}
N_{\l\l'}^{\rm coh}(t,t')=X_{\l\bz}(t)X_{\l'\bz}^{*}(t')+\callO(1-|\a(0)|^2).
\end{align}
Similar to the one-particle Green's function, we expand the 
incoherent contribution in terms of unnormalized exciton states  
\begin{align}
N^{\text{inc}}_{\lambda\lambda'\blQ}
(t, t^{\prime})
&=\bra\Q(t')|\hat{P}^{\rm inc} 
\hat{X}_{\lambda' \blQ}^{\dagger} 
e^{-i \hat{H}( t^{\prime}-t)} 
\hat{X}_{\lambda \blQ}
\hat{P}^{\rm inc}|\Q(t)\ket\nn\\
=& \sum_{\l_1\l_2}\b_{\l_1}(t)\b_{\l_2}^*(t')
\ell(t)\ell^*(t')e^{-iE_{\l_1} t}e^{iE_{{\l_2}}t'}
\sum_{\blQ_1\blQ_2}
\bra \bar{X}_{\l_2\blQ_2}^{\rm inc}|
\hat{X}_{\lambda' \blQ}^{\dagger} 
e^{-i \hat{H}( t^{\prime}-t)} 
\hat{X}_{\lambda \blQ}
|\bar{X}^{\rm inc}_{\l_1\blQ_1}\ket.
\label{App_Ninc}
\end{align}
Considering Eq.~(\ref{Xbar}), the inner product reads
\begin{align}
\bra \bar{X}_{\l_2\blQ_2}^{\rm inc}|
\hat{X}_{\lambda' \blQ}^{\dagger} 
e^{-i \hat{H}( t^{\prime}-t)} 
\hat{X}_{\lambda \blQ}
|\bar{X}^{\rm inc}_{\l_1\blQ_1}\ket
&=\sum_{M,M'=1}^\infty\frac{1}{M!M'!}
\sum_{\substack{\blq_1\ldots\blq_M\\\blq_1'\ldots\blq_{M'}'}}
\d_{\blq_1\ldots +\blq_M,-\blQ_1}
\d_{\blq_1'\ldots +\blq_{M'}',-\blQ_2}
f^*_{\blq_1'}(t') \ldots f^*_{\blq_{M'}'}(t')
f_{\blq_1}(t) \ldots f_{\blq_M}(t)
\nn\\
&\quad\times \bra X_{\l_2\blQ_2}|
\hat{b}_{\blq_{M'}'}\ldots
\hat{b}_{\blq_1'}
\hat{X}_{\lambda' \blQ}^{\dagger} 
e^{-i \hat{H}( t^{\prime}-t)} 
\hat{X}_{\lambda \blQ}
\hat{b}^{\dag}_{\blq_1}\ldots
\hat{b}^{\dag}_{\blq_{M}}
|X_{\l_1\blQ_1}\ket.
\end{align}
After the annihilation of the exciton,
we have a state with only phonons over ground state, 
which is also an eigenstate of $\hat{H}$. Therefore
\begin{align}
e^{-i \hat{H}( t^{\prime}-t)} 
\hat{X}_{\lambda \blQ}
\hat{b}^{\dag}_{\blq_1}\ldots
\hat{b}^{\dag}_{\blq_{M}}
|X_{\l_1\blQ_1}\ket
=\d_{\l\l_1}\d_{\blQ,\blQ_1}
e^{-i(\w_{\blq_1}\ldots+\w_{\blq_M})(t'-t)}
\hat{b}^{\dag}_{\blq_1}\ldots
\hat{b}^{\dag}_{\blq_{M}}
|\F_g\ket.
\end{align}
By following the same steps as outlined in 
Eqs.~(\ref{GF_S_1}--\ref{GF_S_4}), we obtain:  
\begin{align}
\bra \bar{X}_{\l_2\blQ_2}^{\rm inc}|
\hat{X}_{\lambda' \blQ}^{\dagger} 
e^{-i \hat{H}( t^{\prime}-t)} 
\hat{X}_{\lambda \blQ}
|\bar{X}^{\rm inc}_{\l_1\blQ_1}\ket
&=\d_{\l\l_1}\d_{\l'\l_2}
\d_{\blQ,\blQ_1}\d_{\blQ,\blQ_2}
\sum_{M=1}^\infty\frac{1}{M!}
\sum_{{\blq_1\ldots\blq_M}}
\d_{\blq_1\ldots +\blq_M,-\blQ}
\prod_{j=1}^M
\left[f^*_{\blq_j}(t')f_{\blq_j}(t)
e^{-i\w_{\blq_j}(t'-t)}\right]\nn\\
&=\d_{\l\l_1}\d_{\l'\l_2}
\d_{\blQ,\blQ_1}\d_{\blQ,\blQ_2}
S_\blQ(t,t'),
\end{align}
and substitution into Eq.~(\ref{App_Ninc}) yields
\begin{align}
N_{\l\l'\blQ}^{\rm inc}(t,t')
=S_\blQ(t,t')N_{\l\l'}^{\rm coh}(t,t').
\end{align}

\subsection{Many-Body Reduced Density Matrix}
We extend the definition of $\a(t)$ and $\b(t)$ to positive times as
\begin{align}
\a(t>0)=\a(0),\quad\b(t>0)=\b(0)
\end{align}
and the definition of $f_{\blq}(t)$ and $\ell(t)$ at negative times as
\begin{align}
f_{\blq}(t<0)=0,\quad\ell(t<0)=1.	
\end{align}
The many-body reduced density matrix at all time $t$ reads
\begin{align}
\hat{\r}_{\rm el}(t)
=&\Tr_{\rm ph}{\{|\Q(t)\ket\bra\Q(t)|\}}\nn\\
=&
\sum_{\bar{M}=0}^\iif\frac{1}{\bar{M}!}
\sum_{\bar{\blq}_1\ldots\bar{\blq}_{\bar{M}}}
\hat{\mathbb{1}}_{\rm el}
\hat{b}_{\bar{\blq}_{\bar{M}}}\ldots
\hat{b}_{\bar{\blq}_1}
|\Q(t)\ket\bra\Q(t)|
\hat{b}_{\bar{\blq}_1}^{\dag}\ldots
\hat{b}_{\bar{\blq}_{\bar{M}}}^{\dag}
\hat{\mathbb{1}}_{\rm el},
\label{App_dm1}
\end{align}
where we define the identity operator in electronic space as
\begin{align}
\hat{\mathbb{1}}_{\rm el}=
|\F_g\ket\bra\F_g|+\sum_{\l\blQ}
|X_{\l\blQ}\ket\bra X_{\l\blQ}|.
\label{unit_op}
\end{align}
We expand the many body density matrix using the explicit form of 
$|\Q(t)\ket$:
\begin{align}
|\Q(t)\ket\bra\Q(t)|
&=|\a(t)|^2|\F_g\ket\bra\F_g|+
\sum_\l\left(\a^*(t)\b_\l(t)\ell(t)
e^{-iE_\l t}|X_{\l\bz}\ket\bra\F_g|+
\text{H.c.}\right)+
\sum_{\l\blQ}
\left(\a^*(t)\b_\l(t)\ell(t)
e^{-iE_\l t}|\bar{X}^{\rm inc}_{\l\blQ}\ket\bra\F_g|+
\text{H.c.}\right)\nn\\
&\quad+\sum_{\l\l'}\b_{\l}(t)\b_{\l'}^*(t)
|\ell(t)|^2e^{-i(E_\l-E_{\l'})t}
|X_{\l\bz}\ket\bra X_{\l'\bz}|+
\sum_{\l\l'\blQ}
\left(\b_{\l}(t)\b_{\l'}^*(t)
|\ell(t)|^2e^{-i(E_\l-E_{\l'})t}
|\bar{X}^{\rm inc}_{\l\blQ}\ket\bra X_{\l'\bz}|+
\text{H.c.}\right)\nn\\
&\quad+\sum_{\l\l'\blQ\blQ'}
\b_{\l}(t)\b_{\l'}^*(t)
|\ell(t)|^2e^{-i(E_\l-E_{\l'})t}
|\bar{X}^{\rm inc}_{\l\blQ}\ket
\bra \bar{X}^{\rm inc}_{\l'\blQ'}|.
\end{align}
It is easily to verify that
\begin{align}
\Tr_{\rm ph}{\{|\F_g\ket\bra\F_g|\}}
&=|\F_g\ket\bra\F_g|,\\
\Tr_{\rm ph}{\{|X_{\l\bz}\ket\bra\F_g|\}}
&=|X_{\l\bz}\ket\bra\F_g|,\\
\Tr_{\rm ph}{\{|X^{\rm inc}_{\l\blQ}\ket\bra\F_g|\}}
&=\hat{\mathbb{0}}_{\rm el},\\
\Tr_{\rm ph}{\{|X_{\l\bz}\ket\bra X_{\l'\bz}|\}}
&=|X_{\l\bz}\ket\bra X_{\l'\bz}|,\\
\Tr_{\rm ph}{\{|\bar{X}^{\rm inc}_{\l\blQ}\ket\bra X_{\l'\bz}|\}}
&=\hat{\mathbb{0}}_{\rm el}.
\end{align}
As for the pure incoherent contribution, we have
\begin{align}
\Tr_{\rm ph}{\{|\bar{X}^{\rm inc}_{\l\blQ}\ket
\bra \bar{X}^{\rm inc}_{\l'\blQ'}|\}}
=&\sum_{\bar{M}=0}^\iif\frac{1}{\bar{M}!}
\sum_{\bar{\blq}_1\ldots\bar{\blq}_{\bar{M}}}
\hat{\mathbb{1}}_{\rm el}
\hat{b}_{\bar{\blq}_{\bar{M}}}\ldots
\hat{b}_{\bar{\blq}_1}
|\bar{X}^{\rm inc}_{\l\blQ}\ket
\bra \bar{X}^{\rm inc}_{\l'\blQ'}|
\hat{b}_{\bar{\blq}_1}^{\dag}\ldots
\hat{b}_{\bar{\blq}_{\bar{M}}}^{\dag}
\hat{\mathbb{1}}_{\rm el}.
\end{align} 
We expand the state according to Eq.~(\ref{Xbar})
\begin{align}
\hat{\mathbb{1}}_{\rm el}
\hat{b}_{\bar{\blq}_{\bar{M}}}\ldots
\hat{b}_{\bar{\blq}_1}
|\bar{X}^{\rm inc}_{\l\blQ}\ket
=&\sum_{M=1}^\infty
\frac{1}{M!}
\sum_{ \blq_1 \ldots \blq_M}
\delta_{\blq_1\ldots +\blq_M,-\blQ}
f_{\blq_1}(t) \ldots f_{\blq_M}(t)
\hat{\mathbb{1}}_{\rm el}
\hat{b}_{\bar{\blq}_{\bar{M}}}\ldots
\hat{b}_{\bar{\blq}_1}
\hat{b}^{\dag}_{\blq_1}\ldots\hat{b}^{\dag}_{\blq_M}
|{X}_{\l\blQ}\rangle\nn\\
=&\sum_{M=1}^\infty
\frac{1}{M!}
\sum_{ \blq_1 \ldots \blq_M}
\delta_{\blq_1\ldots +\blq_M,-\blQ}
f_{\blq_1}(t) \ldots f_{\blq_M}(t)
\d_{\bar{M}M}\sum_P\prod_{j=1}^M\d_{\bar{\blq}_j,\blq_{P(j)}}
|{X}_{\l\blQ}\rangle\nn\\
=&(1-\d_{\bar{M},0})\d_{\bar{\blq}_1\ldots+\bar{\blq}_{\bar{M}},-\blQ}
f_{\bar{\blq}_1}(t) \ldots f_{\bar{\blq}_{\bar{M}}}(t)
|{X}_{\l\blQ}\rangle,
\end{align} 
where we use relations~(\ref{unit_op}) and (\ref{orth_nor_rel_Xph}) 
to derive the second line. Then
\begin{align}
\Tr_{\rm ph}{\{|\bar{X}^{\rm inc}_{\l\blQ}\ket
\bra \bar{X}^{\rm inc}_{\l'\blQ'}|\}}
=&\d_{\blQ,\blQ'}
\sum_{\bar{M}=1}^\iif\frac{1}{\bar{M}!}
\sum_{\bar{\blq}_1\ldots\bar{\blq}_{\bar{M}}}
\d_{\bar{\blq}_1\ldots+\bar{\blq}_{\bar{M}},-\blQ}
|f_{\bar{\blq}_1}(t)|^2 \ldots |f_{\bar{\blq}_{\bar{M}}}(t)|^2
|{X}_{\l\blQ}\ket
\bra {X}_{\l'\blQ} |\nn\\
=&\d_{\blQ,\blQ'}S_{\blQ}(t)|{X}_{\l\blQ}\ket
\bra {X}_{\l'\blQ} |.
\end{align} 
Substituting this result into Eq.~(\ref{App_dm1}) we find
\begin{align}
\hat{\r}_{\rm el}(t)
&=|\a(t)|^2|\F_g\ket\bra\F_g|+
\sum_\l\left(\a^*(t)\b_\l(t)\ell(t)
e^{-iE_\l t}|X_{\l\bz}\ket\bra\F_g|+
\text{H.c.}\right)\nn\\
&\quad+\sum_{\l\l'}\b_{\l}(t)\b_{\l'}^*(t)
|\ell(t)|^2e^{-i(E_\l-E_{\l'})t}
|X_{\l\bz}\ket\bra X_{\l'\bz}|
\nn\\
&\quad+\sum_{\l\l'\blQ}
\b_{\l}(t)\b_{\l'}^*(t)
|\ell(t)|^2e^{-i(E_\l-E_{\l'})t}
S_\blQ(t)
|{X}_{\l\blQ}\ket
\bra {X}_{\l'\blQ}|.
\end{align}
By identifying the coefficients of $\hat{\r}_{\rm el}$ 
with the excitonic polarization and the exciton density matrix, see 
Eqs.~(\ref{xlcoh},\ref{Xcoh_dm},\ref{Xinc_dm}), we arrive at the 
sought result  
\begin{align}
\hat{\rho}_{\mathrm{el}}(t) &=|\alpha(t)|^2
 |\Phi_g\ket
 \bra\Phi_g|+ \sum_{\lambda\lambda'}N^{\text{coh}}_{\lambda\lambda'}(t)
 |X_{\l\bz}\ket\bra X_{\l'\bz}|\nn\\
 &+\sum_\lambda\left(
 X_{\lambda \bz}(t)
|X_{\l\bz}\ket\bra
\Phi_g|+X_{\lambda \bz}^{\ast}(t)
|\Phi_g\ket\bra
X_{\l\bz}|
\right)
 \nn\\
& 
+\sum_{\lambda\lambda'\blQ}N_{\lambda\lambda'\blQ}^{\text{inc}}(t)
|X_{\lambda\blQ}\ket\bra X_{\lambda'\blQ}|.
\end{align}

\end{widetext}

%\bibliography{reference}% Produces the bibliography via BibTeX.


\end{document}